\documentclass[twocolumn]{aastex62}

\usepackage{txfonts}
\usepackage{color,bm}

\newcommand{\eqref}[1]{(\ref{#1})}

\shorttitle{Atmospheric dynamics and light curves of eccentric-tilted exoplanets}
\shortauthors{Ohno \& Zhang}

\begin{document}

\title{Atmospheres on Nonsynchronized Eccentric-Tilted Exoplanets II: Thermal Light Curves}

\author{Kazumasa Ohno}
\affil{Department of Earth and Planetary Sciences, Tokyo Institute of Technology, Meguro, Tokyo, 152-8551, Japan}

\author{Xi Zhang}
\affil{Department of Earth and Planetary Sciences, University of California Santa Cruz, 1156 High St, Santa Cruz, CA 95064, USA }

\begin{abstract}
Thermal light-curve analysis is a powerful approach to probe the thermal structures of exoplanetary atmospheres, which are greatly influenced by the planetary obliquity and eccentricity.
Here we investigate the thermal light curves of eccentric-tilted exoplanets across various radiative timescales, eccentricities, obliquities, and viewing geometries using results of shallow-water simulations presented in \citet{Ohno&Zhang19}.
We also achieve an analytical theory of the thermal light curve that can explain general trends in the light curves of tilted exoplanets.
For tilted planets in circular orbits, the orbital phase of the flux peak is largely controlled by either the flux from the hot spot projected onto the orbital plane or the pole heated at the summer solstice, depending on the radiative timescale $\tau_{\rm rad}$, planetary day $P_{\rm orb}$, and obliquity $\theta$.
We find that tilted planets potentially produce the flux peak after the secondary eclipse when obliquity is $\theta\ga{90}^{\circ}$ for the hot regime $\tau_{\rm rad}\ll P_{\rm rot}$, or $\theta\ga {18}^{\circ}$ for the cool regime $\tau_{\rm rad}\gg P_{\rm rot}$.
For tilted planets in eccentric orbits, the shape of the light curve is considerably influenced by the heating at the periapse. 
The flux peak occurring after the secondary eclipse can be used to distinguish tilted planets from nontilted planets when the periapse takes place before the secondary eclipse.
Our results could help to constrain exoplanet obliquities in future observations.
\end{abstract}
\keywords{planets and satellites: atmospheres -- planets and satellites: gaseous planets }

%%%%%%%%%%%%%%%%%%%%%
\section{Introduction} \label{sec:intro}

As stated in \citet[][Paper I]{Ohno&Zhang19}, planetary obliquity---the angle between the planet rotation axis and its orbital normal---potentially encapsulates information about planetary climate \citep[e.g.,][]{Williams&Kasting97, Williams&Pollard03,Kane&Torres17} and the formation and evolutionary history of the planet \citep[e.g.,][]{Chambers01,Winn&Holman05,Kokubo&Ida07}.
Thus, retrieving exoplanet obliquities from observations will offer new clues to many important problems on those planets.
To constrain the exoplanet obliquity, a number of observational methods have been proposed to date, for example, oblateness measurement \citep[e.g.,][]{Seager&Hui02}, spin-orbit tomography \citep{Fujii&Kawahara12, Kawahara16}, polarimetry \citep{deKok+11}, and eclipse mapping \citep{Rauscher17}. 
However, no exoplanet obliquity has yet been successfully measured.

Observation of a thermal light curve---a time variation of planetary flux---is a promising way to constrain the exoplanet obliquity.
The light-curve observations probe horizontal temperature distributions on exoplanets, which are strongly associated with atmospheric dynamics \citep[for a recent review, see][]{Parmentier&Crossfield17}.
A number of previous studies have thoroughly investigated the light curves for tidally locked exoplanets with zero obliquity using general circulation models \citep[GCMs; e.g.,][]{Cooper&Showman05,Fortney+06,Showman+09,Kataria+14,Oreshenko+16,Parmentier+16,Komacek+17,Zhang&Showman16,Steinrueck+18,Komacek&Abbot19}.
Several studies have also investigated thermal light curves for nonsynchronized exoplanets.
For example, \citet{Showman+15} examined the light curves of warm and hot Jupiters with various rotation periods and showed that a slower-rotating planet produced a larger amplitude of the light curve for a given stellar irradiation \citep[see also][]{Rauscher&Kempton14}.
\citet{Penn&Vallis17,Penn&Vallis18} also studied the light curves of Earth-like exoplanets with various rotation periods and velocities of the substellar point.
They showed that planets potentially produce the flux peak before or after the secondary eclipse, depending on the rotation period, the substellar velocity, and the gravity wave speed. 
\citet{Kataria+13} investigated the thermal light curves of eccentric hot Jupiters and found that the shape of the light curve highly depends on viewing geometry \citep[see also][]{Langton&Laughlin08,Lewis+10,Lewis+14,Lewis+17}.
Although the above studies have focused on the light curves of nonsynchronized exoplanets, all of them have assumed zero planetary obliquity.

Pioneering study of \citet{Gaidos&Williams04} investigated the infrared light curves of Earth-like planets with nonzero obliquities using an energy balance model and showed that the shape of the light curve is sensitive to the obliquity and the orbital phase of the equinoxes. 
\citet{Langton&Laughlin07} examined the light curves of hot Jupiter with obliquity of 90 deg using a shallow water model. They suggested the shape of the light curves for both nontilted and tilted planets are very similar.
Recently, \citet{Rauscher17} investigated atmospheric circulations on planets with non-zero obliquities and resulting thermal light curves.
%suggested that the seasonal variation occurs when the planetary obliquity is higher than $\sim{30}^{\circ}$.
It was found that the thermal light curve is influenced by not only obliquity but also by the viewing orientation.
In addition, it was also suggested that the peak offset---the orbital phase of the flux peak compared to the secondary eclipse \citep{Parmentier&Crossfield17}---is independent of the planetary obliquity.
However, \citet{Rauscher17} only investigated the light curves of planets in the dynamical regime controlled by diurnally averaged insolation.
As shown in Paper I, horizontal temperature patterns are significantly different at different regimes, which may result in different light curves. 
Moreover, \citet{Rauscher17} only investigated the light curves of planets in a circular orbit.
Because orbital eccentricity is much more difficult to damp than planetary obliquity by the stellar tides during planetary migration \citep{Peale99}, it is expected that tilted \footnote{In this study, ''tilted'' does not mean the inclined orbital plane, namely, a nonzero orbital inclination. Here "tilted" means that the planet rotation axis is misaligned to its orbital normal.} planets are also likely to have nonzero eccentricities, another factor influencing the atmospheric dynamics and transit light curves.

%The purpose of this study is to investigate the atmospheric dynamics on generic eccentric-tilted exoplanets (ET planets hereafter).
In this study, we investigate the thermal light curves of eccentric-tilted exoplanets (ET planets) across a range of relevant parameters: radiative timescale, obliquity, eccentricity, and the viewing geometry.
Then we discuss how to potentially infer the obliquity from the observations.
%We also develop a general analytical theory for the thermal light curve for arbitrary obliquity, radiative timescale, and viewing geometry.
The organization of this paper is as follows.
We overview the dynamical regimes of ET planets in Section \ref{sec:regimes}.
We will present a general analytical theory for the thermal light curve for arbitrary obliquity, radiative timescale, and viewing geometry in Section \ref{sec:lcurve_analy}.
We will show the synthetic thermal light curves of ET planets and discuss the observable signature of nonzero obliquity in Section \ref{sec:lightcurve}.
We summarize this paper in Section \ref{sec:summary}.

\section{Overview of the Dynamical Regimes}\label{sec:regimes}
%%%%%%%%%%%%%%%%%%%%%%
\begin{figure*}[t]
\centering
\includegraphics[clip, width=\hsize]{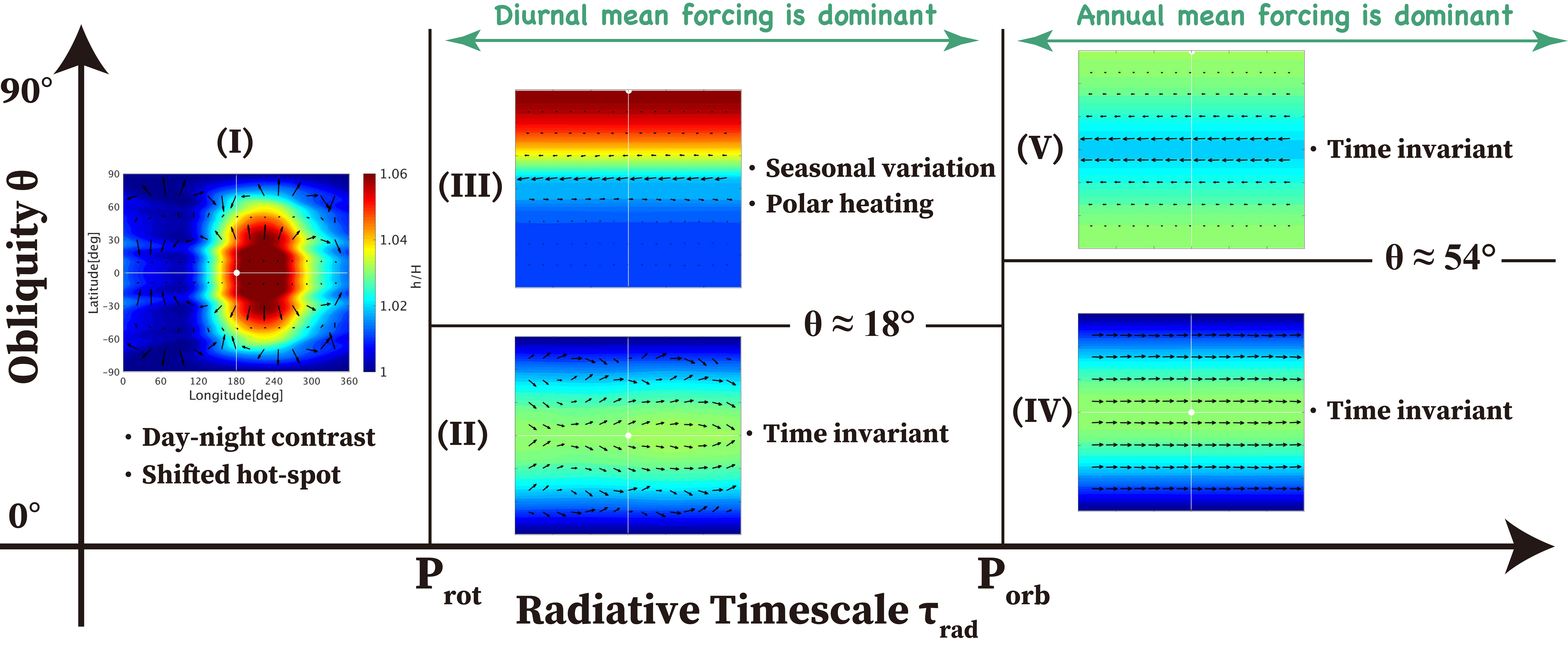}
\caption{Schematic diagram of dynamical regimes for ET planets modified from those in Paper I with a specific focus on temperature patterns. Each panel shows the snapshot of the height field (color scales) and flow pattern (arrows) taken from Paper I. 
}
\label{fig:regimes}
\end{figure*}
%%%%%%%%%%%%%%%%%%%%%%
Here we first briefly summarize the dynamical regimes of ET planets demarcated in Paper I, where we simulated atmospheric circulations on ET planets using a one-and-a-half-layer shallow-water model.
The model results will be used to calculate the synthetic thermal light curves in this paper. 
In paper I, it was shown that the dynamical patterns can be demarcated into five regimes using the radiative timescale and obliquity, as summarized in Figure \ref{fig:regimes}.
For a strongly illuminated planet where the radiative timescale $\tau_{\rm rad}$ is much shorter than the planetary day $P_{\rm rot}$ (regime (I) in Figure \ref{fig:regimes}), the atmospheric circulation is controlled by instantaneous heating patterns.
The atmosphere also exhibits a strong day--night temperature contrast and the eastward-shifted hot spot from the substellar point, originating from a time delay of the atmospheric response to the heating.
When the radiative timescale $\tau_{\rm rad}$ is longer than the planetary day $P_{\rm rot}$ but shorter than the planetary year $P_{\rm orb}$ (regimes (II) and (III) in Figure \ref{fig:regimes}), the circulation is controlled by the diurnally averaged insolation.
In this regime, the atmosphere experiences a significant seasonal variation and an intense heating in the polar regions if the obliquity $\theta$ is higher than the critical value ($\approx{18}^{\circ}$; see Section 2 of Paper I).
For a cold planet where the radiative timescale $\tau_{\rm rad}$ is much longer than the planetary year $P_{\rm orb}$ (regimes (IV) and (V) in Figure \ref{fig:regimes}), the circulation is eventually controlled by the annually averaged insolation.
In this regime, seasonal variations disappear, and the circulation pattern is nearly time-invariant throughout the planetary orbit, although the temperature and flow patterns are different for $\theta\la{54}^{\circ}$ and $\theta\ga{54}^{\circ}$.
Planets with retrograde rotations (i.e., $\theta>{90}^{\circ}$) behave similarly to those with ${180}^{\circ}-\theta$ as long as the planetary day is much shorter than the planetary year, which is true for the solar system planets, except for Venus.
We showed that the regime classification is also applicable to planets in eccentric orbits (for more details, see Paper I).

One can expect several implications of the dynamical regimes on the resulting light curves.
For planets with short radiative timescales ($\tau_{\rm rad}\ll P_{\rm rot}$), the light curve has a large amplitude because of the strong day--night contrast.
The peak offset of the light curve would be controlled by the hot spot shifted from the substellar point, as well as that for a close-in planet.
For an intermediate regime ($P_{\rm rot}\ll \tau_{\rm rad}\ll P_{\rm orb}$), a planets with a large obliquity ($\theta\ga{18}^{\circ}$) would produce the light curve with a large amplitude because of a significant seasonality.
For planets with very long radiative timescales ($\tau_{\rm rad}\gg P_{\rm orb}$), the light curve would be almost flat, since the temperature pattern is nearly time invariable throughout the planet orbit.
We will demonstrate these behaviors using both analytical models and numerical simulations in subsequent sections.

%%%%%%%%%%%%%%%%%%%
\section{Analytical Theory of Thermal Light Curves for Tilting Planets}\label{sec:lcurve_analy}
In this section, we present an analytical model of thermal light curves for tilted planets.
Previous studies also derived analytical models of light curves for tidally locked planets \citep{Zhang&Showman16,Hammond&Pierrehumbert18}, eccentric planets \citep[e.g.,][]{Cowan&Agol11}, and nonsynchronized planets \citep{Penn&Vallis17}; however, all of them assumed planets with zero obliquities.
Here we construct the analytical theory that predicts the orbital phase of the flux peak $f_{\rm peak}$ and the amplitude of the light curves as a function of arbitrary radiative timescale, obliquity, and the viewing geometry.
The complete derivation of the theory is
summarized in Appendices \ref{sec:appendix1}--\ref{sec:appendix4}. 
The most important assumption in our theory is that the emergent flux from the planet mainly consists of the flux from the hot spot shifted from the substellar point and the heated pole (see Figure \ref{fig:regimes}).
The thermal light curve thus originates from the time evolution of the hot spot and the heated pole only (Equation \ref{eq:delta} in Appendix \ref{sec:appendix4}).
The analytical theory well explains the numerical results (Section \ref{sec:lcurve_T}) and offers insights into the basic behaviors of the light curves of tilted planets.

\begin{figure*}[t]
\centering
\includegraphics[clip, width=\hsize]{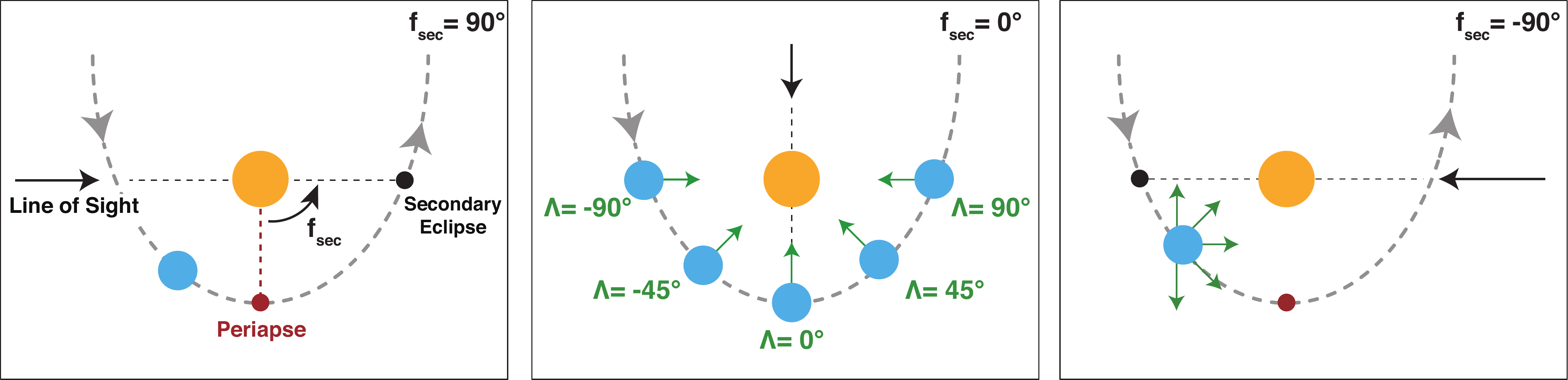}
\caption{ Schematic illustration of the geometry for ET planets. The gray dotted curves show the orbit trajectory, the red dots represent the orbital phases of the periapse, the green arrows represent the planetary rotation axis projected on the orbital plane, and the black dots are the orbital phases of the secondary eclipse. From left to right, the secondary eclipse takes place at $f={90}^{\circ}$, $0^{\circ}$, and $-{90}^{\circ}$, sequentially. The difference of the light-curve shape is also induced by $\Lambda$, which is defined as the angle from orbital phase between the secondary eclipse $f_{\rm sec}$ and the northern summer solstice $f_{\rm sol}$ (see Section \ref{sec:lightcurve}). 
}
\label{fig:geometry}
\end{figure*}

For convenience, we introduce the parameter $\Lambda$, defined as phase angle between the secondary eclipse $f_{\rm sec}$ and the northern summer solstice $f_{\rm sol}$ (see Figure \ref{fig:geometry}),
\begin{equation}\label{eq:lambda}
\Lambda \equiv f_{\rm sol}-f_{\rm sec}.
\end{equation}
Here $\Lambda$ characterizes the viewing geometry; for example, the northern summer solstice takes place before (after) the secondary eclipse for negative (positive) $\Lambda$. 
Note that one can assume $f_{\rm sec}=0$ for planets in circular orbits.
For a given $\Lambda$, the emergent flux $F$ can be calculated as a function of orbital phase from the secondary eclipse $f$ (Equation \eqref{eq:Analytic_Flux} in Appendix \ref{sec:appendix4}):
\begin{equation}\label{eq:Flux_analytic}
F(f)=\pi H+\Delta h C_{\rm t}(\theta,\varphi,\Lambda,\psi)\cos{[f-\Lambda+\varphi_{\rm peak}(\theta,\varphi,\Lambda,\psi)]}.
\end{equation}
Here we adopt the shallow-water framework in Paper I.
Here $H$ is the mean atmospheric height on the nightside (i.e., $gH$ is the mean geopotential on the nightside), $\Delta h$ is the difference of the equilibrium height between the substellar point and the nightside (see Paper I), $\psi$ is the dimensionless parameter defined as (Equation \eqref{eq:psi} in Appendix \ref{sec:appendix3})
\begin{equation}
\psi \equiv \frac{P_{\rm orb}}{2\pi \tau_{\rm rad}},
\end{equation}
and $C_{\rm t}$ and $\varphi_{\rm peak}$ are the parameters controlling the amplitude and flux peak phase of the light curve, given by Equations \eqref{eq:Ct} and \eqref{eq:appendix_full} in Appendix \ref{sec:appendix4}.
The parameters depend on the phase shift of the hot spot from the substellar point on the equatorial plane $\varphi$, which can be evaluated by solving (see Appendix \ref{sec:appendix2})
\begin{equation}\label{eq:original_offset}
\xi \sin{\varphi}-\cos{\varphi}+\xi \frac{e^{(3/2)\pi \xi}+e^{(1/2)\pi \xi}}{e^{2\pi \xi}-1}\exp{(-\xi \varphi)}=0,
\end{equation}
where $\xi$ is the nondimensional parameter, defined as
\begin{equation}
\xi \equiv \frac{P_{\rm rot}}{2\pi \tau_{\rm rad}}\left(1- \frac{gH}{v_{\rm ss}^2}\right)^{-1},
\end{equation}
where $g$ is the surface gravity, and $v_{\rm ss}=2\pi R_{\rm p}/P_{\rm rot}$ is the substellar velocity.
In the case of $\xi \gg 1$, the phase shift is simply $\varphi\approx {\rm tan}^{-1}(1/\xi)$.
Equation \eqref{eq:Flux_analytic} gives the orbital phase of the flux peak $f_{\rm peak}$ as
\begin{equation}
f_{\rm peak}=\Lambda - \varphi_{\rm peak}.
\end{equation}
We note that the theory is basically applicable to planets in circular orbits.
For eccentric orbits, because the true anomaly is not a linear function of time, one needs to perform a numerical integration to evaluate the emergent flux, and therefore no explicit close form is presented here (but it is still a predictive theory from first principles).
We only focus on the analytical cases for circular orbits.

%%%%%%%%%%%%%%%%%%%%%%
%%%%%%%%%%%%%%%%%%%%%%

We present the limiting behavior of the analytical theory to clarify what mechanism is controlling the peak offset for tilted planets.
In the limit of a short radiative timescale ($\tau_{\rm rad}\ll P_{\rm rot}$), the shape of the light curve is mainly dominated by the flux from the shifted hot spot.
This is similar to cases for close-in synchronized planets, but the problem is more complicated because the equatorial plane on a tilted planet is misaligned with the orbital plane on which the subobserver point moves.
When the flux is only contributed by the shifted hot spot, the angle $\varphi_{\rm peak}$ is expressed as (see Appendix \ref{sec:appendix1} for the derivation)
\begin{equation}\label{eq:limit_nopole}
\varphi_{\rm peak}={\rm tan}^{-1}\left[ \frac{\cos{\varphi}\sin{\Lambda}+\sin{\varphi}\cos{\theta}\cos{\Lambda}}{({\rm sin}^{2}\theta+\cos{\varphi}~{\rm cos}^{2}\theta)\cos{\Lambda}-\sin{\varphi}\cos{\theta}\sin{\Lambda}}\right].
\end{equation}
When the original phase shift on the equatorial plane $\varphi$ is sufficiently small, which may be valid in the limit of a short radiative timescale and weak zonal flow, the equation is approximated by (Equation \eqref{eq:appendix3} in Appendix \ref{sec:appendix1})
\begin{equation}
\varphi_{\rm peak}\approx \Lambda +\varphi \cos{\theta}.
\end{equation}
Therefore, the orbital phase of the flux peak is given by 
\begin{equation}\label{eq:limit1}
f_{\rm peak}\approx -\varphi \cos{\theta}.
\end{equation} 
Equation \eqref{eq:limit1} can be interpreted as that the shape of the light curve is controlled by the "projected hot spot" onto the orbital plane in the limit of a short radiative timescale.
It is worth noting that the peak offset only depends on the original phase shift and obliquity in this regime.

The shifted hot spot hardly contributes on the total emergent flux as the radiative timescale increases because the height field becomes more homogenized in longitude (see Figure \ref{fig:regimes}).
In the limit of a long radiative timescale ($\tau_{\rm rad}\gg P_{\rm rot}$), the emergent flux is dominated by the flux from the polar region, which is strongly heated at around the solstice.
When the light curve is only dominated by the flux from the polar region, the phase of the flux peak is equivalent to the phase at which the height field in the pole is maximized, which is approximately given by (see Appendix \ref{sec:appendix3} for the derivation)
\begin{equation}\label{eq:limit2}
f_{\rm peak}\approx \Lambda + {\rm tan}^{-1}\left( \frac{2\pi \tau_{\rm rad}}{P_{\rm orb}} \right),
\end{equation}
where we have assumed that the orbital period is much longer than the radiative timescale ($\psi \gg1$).
When the radiative timescale is much longer than the orbital period ($\psi \ll 1$), the emergent flux no longer shows a time variation because the height fields are nearly constant throughout the planetary orbit (see also Equation \eqref{eq:height_pole_annual} in Appendix \ref{sec:appendix3}).
Equation \eqref{eq:limit2} indicates that the orbital phase of the flux peak is determined by the time lag behind the ''seasonal polar heating" from the solstice.
In the next section, we demonstrate that the ''projected hot spot" and ''seasonal polar heating" control the shape of the light curves of tilted exoplanets using the numerical calculations.

%%%%%%%%%%%%%%%%%%%
\section{Synthetic Thermal Light Curves}\label{sec:lightcurve}
We studied the thermal light curves of ET planets using the results of shallow-water simulations presented in Paper I.
The thermal light curve can be calculated as a time variation of the emergent flux $F$ from the hemisphere facing to the observer.
Following previous studies using a shallow-water model \citep{Zhang&Showman14,Penn&Vallis17}, we calculate the synthetic thermal light curves by integrating the height fields simulated in Paper I over the visible hemisphere as a proxy of the emergent flux,
\begin{equation}\label{eq:lcurve}
F(t)=\int^{2\pi}_{0}\int^{\pi/2}_{-\pi/2}h(t)(\mathbf{r}\cdot \mathbf{r}_{\rm obs})\mathcal{H}(\mathbf{r}\cdot \mathbf{r}_{\rm obs})\cos{\phi}d\phi d\lambda,
\end{equation}
where $h$ is the atmospheric height from the shallow-water model, $\phi$ is the latitude, $\lambda$ is the longitude, and $\mathbf{r}_{\rm obs}$ is the point vector from the distant planet to the observer referred to as the subobserver point \citep{Rauscher17}.
The subobserver latitude $\phi_{\rm obs}$ is related to $\Lambda$ as
\begin{equation}\label{eq:phi_obs}
    \sin{\phi_{\rm obs}}=\sin{\theta}\cos{\Lambda}.
\end{equation}
Here $\mathcal{H}(x)$ is the Heaviside step function accounting for the fact that the hemisphere not facing to the observer is invisible, defined as
\begin{equation}\label{eq:Heaviside}
    \mathcal{H}(x) =  
\left\{
\begin{array}{ll}
      1  & \text{when $x \ge 0$} \\
      0 & \text{when $x < 0$}.
         \end{array}
\right.
\end{equation}
We note here that the light curve calculated in Equation \eqref{eq:lcurve} does not take into account the telescope integration time in real observations.
If the integration time is long, the light curve will be essentially smeared out in a certain time window.

\subsection{Tilted Planets in Circular Orbits}\label{sec:lcurve_T}
\begin{figure*}[t]
\centering
\includegraphics[clip, width=\hsize]{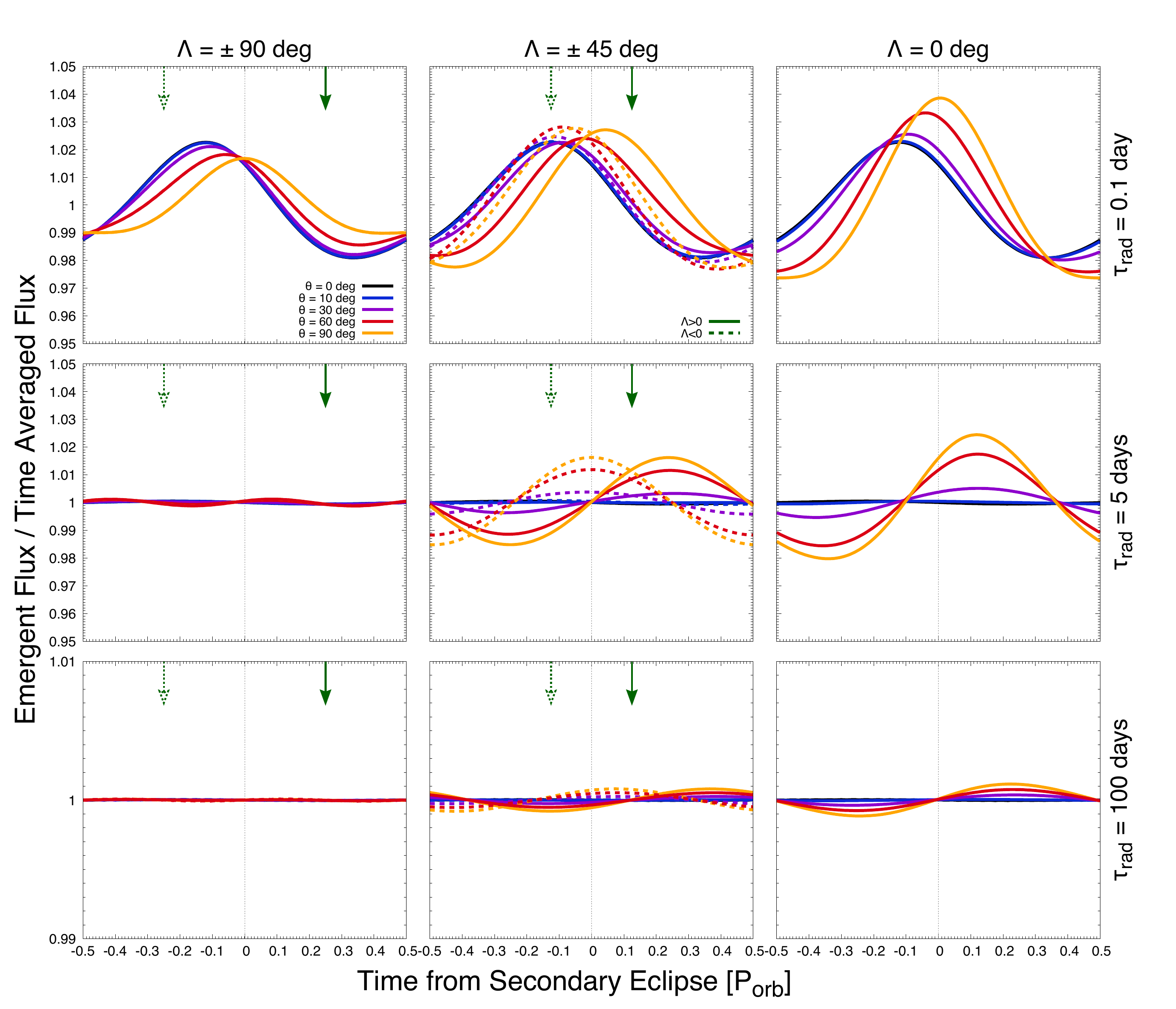}
\caption{
Thermal light curves of planets in circular orbits for different $\tau_{\rm rad}$, $\theta$, and $\Lambda$. The vertical and horizontal axes are the emergent flux normalized by the time-averaged flux and the time from the secondary eclipse, respectively. From top to bottom, each the rows show the cases of $\tau_{\rm rad}=0.1$, $5$, and $100~{\rm days}$, respectively. The columns, from left to right, show the cases of $\Lambda=\pm {90}^{\circ}$, $\pm {45}^{\circ}$, and $0^{\circ}$, respectively.
The black, navy, purple, red, and orange lines show the light curves for $\theta=0^{\circ}$, ${10}^{\circ}$, ${30}^{\circ}$, ${60}^{\circ}$, and ${90}^{\circ}$, respectively.
The vertical dotted lines denote the time of the secondary eclipse.
For the left and middle columns, the solid lines are the light curves of planets whose northern summer solstice takes place before the secondary eclipse ($\Lambda<0$), while the dotted lines are for the planets with summer solstice after the secondary eclipse ($\Lambda>0$).
The green filled and open arrows denote the time of summer solstice that occurs before ($\Lambda<0$) and after ($\Lambda>0$) the secondary eclipse, respectively.
}
\label{fig:lcurve_circ}
\end{figure*}
%%%%%%%%%%%%%%%%%%
\begin{figure*}[t]
\centering
\includegraphics[clip, width=0.9\hsize]{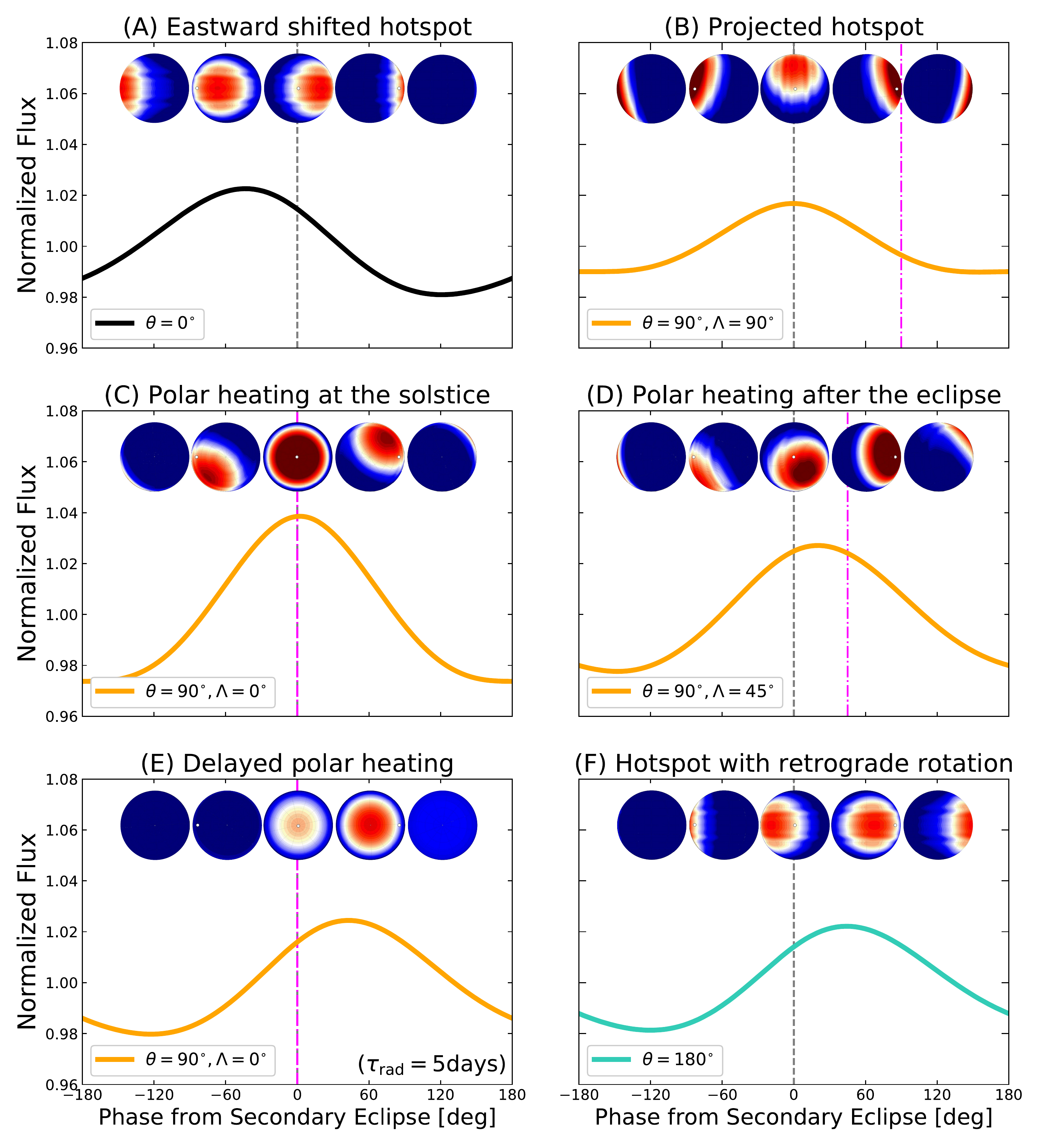}
\caption{
Typical shapes of light curves with height fields on the visible hemisphere. The vertical axis is the same as in Figure \ref{fig:lcurve_circ}, and the horizontal axis is the orbital phase from the secondary eclipse.
The radiative timescale is $\tau_{\rm rad}=0.1~{\rm day}$ for panels (A)--(D) and (F) and $\tau_{\rm rad}=5~{\rm days}$ for panel (E). 
The gray dotted and pink dash-dotted lines denote the phase of secondary eclipse and the northern summer solstice, respectively.
}
\label{fig:lcurve_map}
\end{figure*}
%%%%%%%%%%%%%%%%%%%%

%%%%%%%%%%%%%%%%%%
\begin{figure*}[t]
\centering
\includegraphics[clip, width=\hsize]{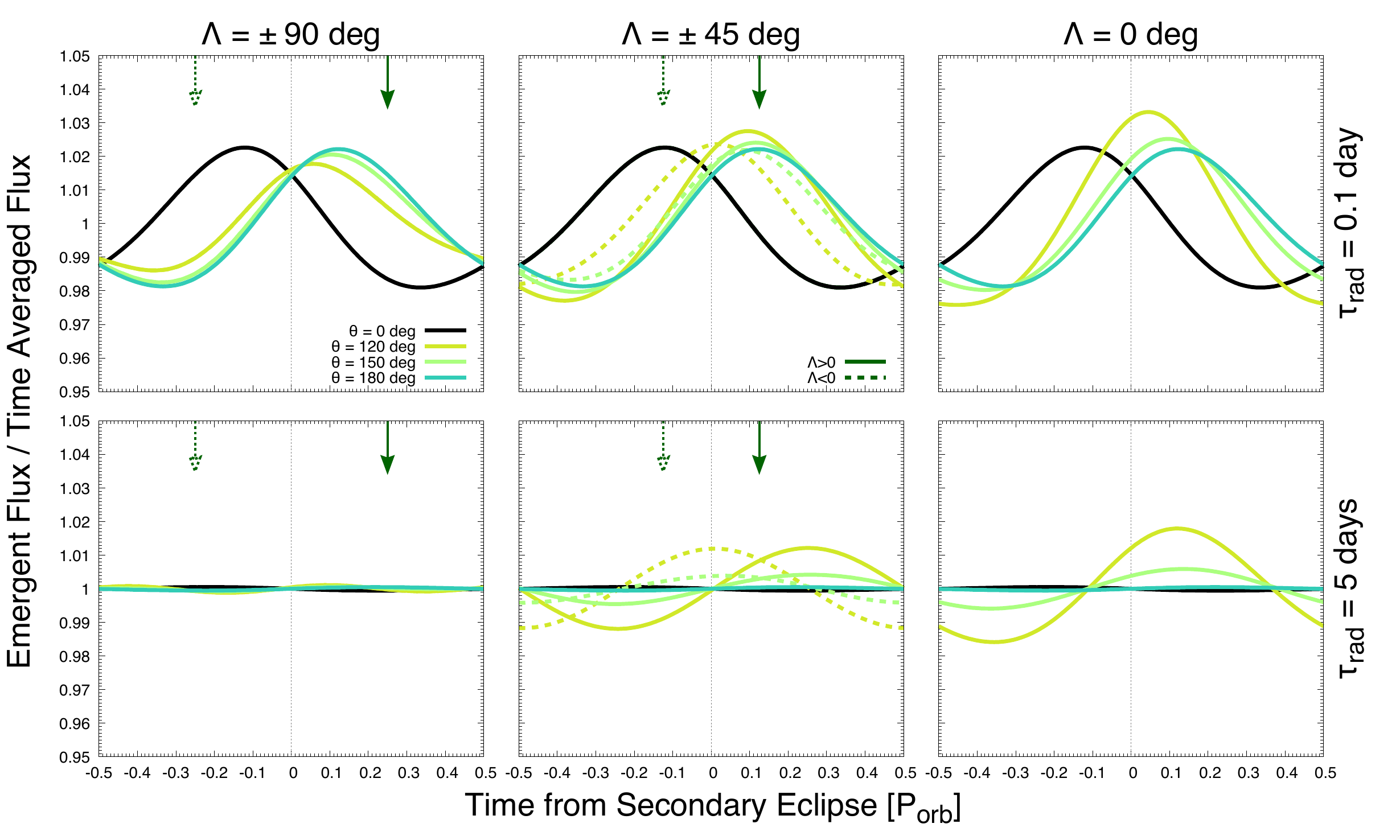}
\caption{
Same as Figure \ref{fig:lcurve_circ} but for planets with retrograde rotation (i.e., $\theta>{90}^{\circ}$). 
The top and bottom rows show the cases of $\tau_{\rm rad}=0.1$ and $5~{\rm days}$, respectively. 
The black, lemon, light green, and peacock green lines show the light curves for $\theta=0^{\circ}$, ${120}^{\circ}$, ${150}^{\circ}$, and ${180}^{\circ}$, respectively.
}
\label{fig:lcurve_retro}
\end{figure*}
%%%%%%%%%%%%%%%%%%%%

First, we show the thermal light curves of tilted planets in a circular orbit for different $\tau_{\rm rad}$, $\theta$, and $\Lambda$ (Figure \ref{fig:lcurve_circ}).
For circular orbits, one can assume $f_{\rm sec}=0^{\circ}$.
We also plot the light curves of nontilted planets for comparison.
Following the convention, the flux peak occurring before the secondary eclipse is referred to a {\it positive} peak offset, while the peak occurring after the secondary eclipse is referred to a {\it negative} peak offset \citep[][]{Parmentier&Crossfield17}.

For nontilted planets, the light curves always show the positive peak offsets when the radiative timescale is short (top row of Figure \ref{fig:lcurve_circ}).
To better understand how the planet looks like at each orbital phase, we show the height fields on the visible hemisphere on top of the light curves (Figure \ref{fig:lcurve_map}).
As seen in panel (A) of Figure \ref{fig:lcurve_map}, the positive peak offset is caused by the eastward shift of the hot spot from the substellar point due to a time delay of the height field in response to stellar irradiation.
As the radiative timescale increases, the height field is more homogenized in longitude (see Figure \ref{fig:regimes}).
Therefore, as the radiative timescale increases, the light-curve amplitude is weaker, and the light curve eventually becomes flat, as shown in the middle and bottom rows of Figure \ref{fig:lcurve_circ}.
These behaviors are consistent with the synthetic light curves for nonsynchronized planets in a circular orbit in previous studies \citep{Showman+15,Penn&Vallis17,Penn&Vallis18,Rauscher17}.

For planets with nonzero obliquities, as shown in Figure \ref{fig:lcurve_circ}, the behaviors of the thermal light curves are very complex, depending on the radiative timescale, planetary obliquity, and the viewing geometry.
These complex behaviors are mainly caused by the geometric effect of the "projected hot spot" and the effect of the "seasonal polar heating" argued in Section \ref{sec:lcurve_analy}.
The former effect is responsible for light curves in regime (I)---those with short radiative timescales ($\tau_{\rm rad}\ll P_{\rm rot}$).
The latter effect is mainly responsible for regime (III)---those with long radiative timescales ($\tau_{\rm rad}\gg P_{\rm rot}$) and large obliquities ($\ga {18}^{\circ}$).
For planets with very weak seasonality in regimes (II), (IV), and (V), the light curves are almost flat because the height fields are nearly constant throughout the planetary orbit (Figure \ref{fig:lcurve_circ}).
%{\bf We interpret each radiative timescale regime in detail from below.}

In the case of $\tau_{\rm rad}=0.1~{\rm day}$ (top row in Figure \ref{fig:lcurve_circ}), both the amplitude and peak offset of the light curves appreciably vary with obliquity because of the aforementioned geometrical effect.
Interestingly, the dependence of obliquity on the amplitude is different for different viewing geometry characterized by $\Lambda$.
For $\Lambda=\pm {90}^{\circ}$ geometry, in which the observer is facing the equator at the secondary eclipse (see Figure \ref{fig:geometry}), the amplitude of the light curve decreases with increasing obliquity (top left panel of Figure \ref{fig:lcurve_circ}).
On the other hand, for $\Lambda={0}^{\circ}$ geometry, in which the solstice occurs at the secondary eclipse and the observer views the polar region then, the amplitude increases with increasing obliquity (top right panel of Figure \ref{fig:lcurve_circ}).
According to the analytical light curve, the amplitude of the light curve is scaled by $C_{\rm t}(\theta,\varphi,\Lambda,\psi)$ (Section \ref{sec:lcurve_analy}).
In the limit of short radiative timescales, the amplitude factor $C_{\rm t}$ can be approximated by $C(\theta,\varphi,\Lambda)\cos{\varphi}$, where $C$ is given by Equation \eqref{eq:amplitude} in Appendix \ref{sec:appendix1}.
The prefactor $C(\theta,\varphi,\Lambda)$ for $\Lambda={90}^{\circ}$ is given by (Equation \eqref{eq:projection90} in Appendix \ref{sec:appendix1})
\begin{equation}
C|_{\Lambda={90}^{\circ}}=\sqrt{{\rm cos}^{2}\varphi+{\rm sin}^{2}\varphi~{\rm cos}^{2}\theta}.
\end{equation}
Therefore, the amplitude decreases with increasing obliquity for $\Lambda={90}^{\circ}$, in agreement with Figure \ref{fig:lcurve_circ}.
This is qualitatively due to the fact that the projected maximum emission flux of the hot spot along the line of the sight to the observer will be smaller if the obliquity is higher in this geometry.
On the other hand, the prefactor $C$ for $\Lambda=0^{\circ}$ is given by (Equation \eqref{eq:projection0} in Appendix \ref{sec:appendix1})
\begin{equation}
C|_{\Lambda={0}^{\circ}}=\sqrt{({\rm sin}^{2}\theta+\cos{\varphi}~{\rm cos}^{2}\theta)^2+{\rm sin}^{2}\varphi~{\rm cos}^{2}\theta}.
\end{equation}
Here the prefactor $C$ is nearly unity for all obliquities as long as the phase shift of the hot spot is small, while Figure \ref{fig:lcurve_circ} indicates that the amplitude increases with increasing obliquity.
This is due to the fact that the actual light curve is obtained from the disk-integrated flux.
The hot spot is less smoothed out at the polar region but more smoothed out at the equator, as seen in panels (A) and (C) in Figure \ref{fig:lcurve_map}, which induces the higher disk-integrated flux from the polar region than that from the equator.

For a short radiative timescale ($\tau_{\rm rad}\ll P_{\rm rot}$), the phase (or time) of the peak in the light curve approaches the secondary eclipse as obliquity increases for $\Lambda=\pm{90}^{\circ}$ and $0^{\circ}$.
According to Equation \eqref{eq:limit1}, the phase of the flux peak approaches the secondary eclipse as the obliquity approaches $\theta={90}^{\circ}$, which is consistent with the trends of the peak offset.
This is qualitatively originated from the fact that the equatorial plane and the orbital planes will be more and more misaligned with each other as the obliquity increases.
For example, in the $\Lambda={90}^{\circ}$ and $\theta={90}^{\circ}$ case (panel (B) in Figure \ref{fig:lcurve_map}), the equatorial plane is essentially perpendicular to the orbital plane; thus, the observer will always see the flux peak from the projected hot spot occurring right at the secondary eclipse.
On the other hand, the peak in the light curve moves toward the phase before and after the secondary eclipse for $\Lambda=-{45}^{\circ}$ and ${45}^{\circ}$, respectively.
This phase shift from the secondary eclipse is caused by the flux from the polar region that enhances the total emergent flux around the solstice when the obliquity is high.
For example, in the $\Lambda={45}^{\circ}$ and $\theta={90}^{\circ}$ case, in which the solstice takes place after the secondary eclipse (panel (D) in Figure \ref{fig:lcurve_map}), the polar region undergoes a strong heating around the solstice, and the peak in the light curve also occurs near there.
Note that the effect of polar heating is not responsible for $\Lambda=\pm{90}^{\circ}$ because the polar region is no longer visible to the observer in that geometry.

In the case of $\tau_{\rm rad}=5~{\rm days}$ (middle row of Figure \ref{fig:lcurve_circ}), the shapes of the light curves are significantly different from those of $\tau_{\rm rad}=0.1~{\rm day}$.
For planets with obliquity smaller than ${18}^{\circ}$ (regime II), the light curves look nearly flat.
The phase offset, if there is one, is shifted before the secondary eclipse.
In this regime, because of the weak seasonality, the peak offset is still determined by the projected hot spot rather than the seasonal polar heating.
On the other hand, the light curves of highly tilted planets exhibit noticeable peaks.
This is caused by the polar heating occurring at around the solstice, which produces a strong emergent flux (see Figure \ref{fig:regimes}).
Since the polar heating occurs if the obliquity is higher than ${18}^{\circ}$ in this regime (Section \ref{sec:regimes}), a planet with $\theta > {18}^{\circ}$ potentially exhibits a flux variation in the light curve.
In this regime, since there is a significant time lag in the heating in the polar region (panel (E) in Figure \ref{fig:lcurve_map}), the flux peak occurs after the solstice, as seen in the light curves for $\Lambda=\pm {45}^{\circ}$ and ${90}^{\circ}$ (middle and right panels in Figure \ref{fig:lcurve_circ}).
This can also be known from Equation \eqref{eq:limit2}, which shows the phase of the flux peak $f_{\rm peak}\approx \Lambda+2\pi \tau_{\rm rad}/P_{\rm orb}$.
The light curves behave nearly flat in the geometry with $\Lambda=\pm{90}^{\circ}$ (left panel in Figure \ref{fig:lcurve_circ}) in which the subobserver point is at the equator and the flux from the poles is negligible.
For $\Lambda=\pm{90}^{\circ}$, highly tilted planets produce double flux peaks in the light curves. 
This is caused by the flux from the equatorial region heated at the vernal and autumn equinoxes. 
The bimodal light curve for this specific geometry can also be seen in previous studies \citep{Gaidos&Williams04,Rauscher17}.

The remarkable feature is that a tilted planet produces a flux peak after the secondary eclipse in some cases, for example, the light curves for $\Lambda=0$ and $\theta \geq {30}^{\circ}$ (middle right panel of Figure  \ref{fig:lcurve_circ}).
This originates from the fact that the phase of the flux peak highly depends on the orbital phase of the solstice when the high latitudes are illuminated.
The middle row of Figure \ref{fig:lcurve_circ} shows that the flux peak could occur after the secondary eclipse for $\theta={30}^{\circ}$, ${60}^{\circ}$, and ${90}^{\circ}$.
Note that $\Lambda>0$ corresponds to the northern summer solstice occurring after the secondary eclipse (see Figure \ref{fig:geometry}).
Because there is also a time lag in the polar heating, the maximum temperature tends to occur after the solstice and the secondary eclipse, as seen in Figure \ref{fig:lcurve_circ}.
These behaviors are in agreement with \citet{Rauscher17} in which the peaks of the light curves tend to occur after the secondary eclipse.
Therefore, one can infer that the obliquity is higher than about ${18}^{\circ}$ if the negative peak offset is found for planets with a long radiative timescale, as classified into regimes (II) and (III).
However, it would be difficult to identify the exact value of the obliquity only from the peak offset because  the peak offset is controlled by solstice phase $\Lambda$ and independent of obliquity (Equation \ref{eq:limit2}).

In the case of $\tau_{\rm rad}=100~{\rm days}$ (bottom row of Figure \ref{fig:lcurve_circ}), the light curves are almost flat for all obliquities.
The amplitudes of the light curves are at least an order of magnitude smaller than the light curves for $\tau_{\rm rad}=5~{\rm days}$.
This is simply because, in regimes (IV) and (V), the height fields are controlled by the annual mean insolation and do not change throughout the planet orbit (Section \ref{sec:regimes}).

%%%%%%%%%%%%%%%%%%%
\begin{figure*}[t]
\centering
\includegraphics[clip, width=\hsize]{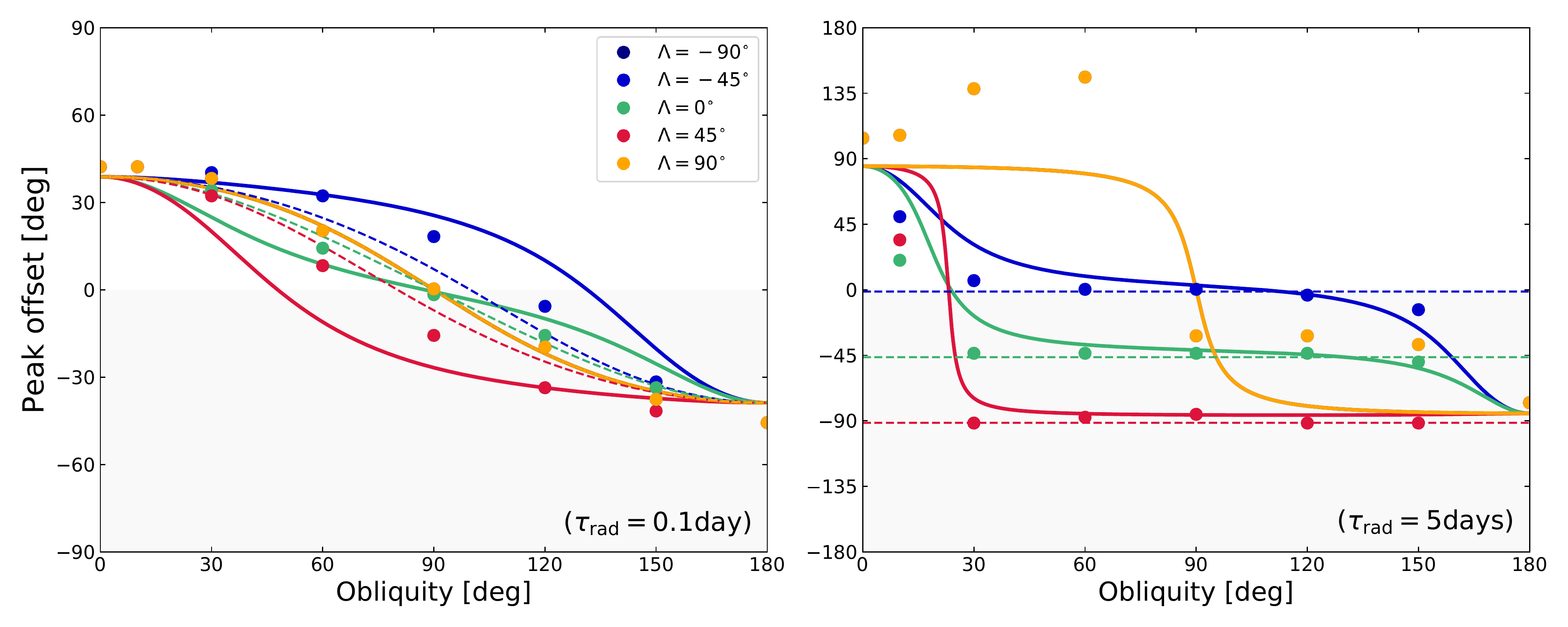}
\caption{Peak offset of light curves of tilted planets as a function of obliquity. 
The left and right panels show the peak offsets for $\tau_{\rm rad}=0.1$ and $5~{\rm days}$, respectively. 
The horizontal axis is the planetary obliquity, and the vertical axis is the peak offset defined as $f_{\rm sec}-f_{\rm peak}=-f_{\rm peak}$.
The dots show the peak offsets calculated by our 2D simulations for different $\Lambda$. 
The solid lines show the peak offset predicted by a universal analytical solution (Equation \eqref{eq:appendix_full} in Appendix \ref{sec:appendix4}) where the original phase shift of the hot spot $\varphi$ is obtained by solving Equation \eqref{eq:original_offset}.
The dotted lines in the left and right panels denote the peak offset derived from the formula assuming the flux from only the projected hot spot (Equation \ref{eq:limit_nopole}) and only the pole (Equation \ref{eq:limit2}), respectively.
When the light curve has several peaks, as seen the case of $\tau_{\rm rad}=5~{\rm days}$ and $\Lambda={90}^{\circ}$(Figure \ref{fig:lcurve_circ}), we measure the peak offset using the largest peak.
}
\label{fig:appendix}
\end{figure*}
%%%%%%%%%%%%%%%%%%%

We also show the light curves for planets with retrograde rotations (i.e., $\theta>{90}^{\circ}$) in Figure \ref{fig:lcurve_retro}.
For $\tau_{\rm rad}=0.1~{\rm day}$, the peak of the light curve occurs significantly after the secondary eclipse as the obliquity approaches $\theta={180}^{\circ}$.
This is again caused by the aforementioned geometrical effect.
The eastward displacement of the hot spot on the equatorial plane is still present for $\theta>{90}^{\circ}$ (see Paper I).
Because the hot spot projected on the orbital plane effectively shifts westward for the observer when $\theta>{90}^{\circ}$ (see Equation \eqref{eq:limit1} and panel (F) of Figure \ref{fig:lcurve_map}), the flux peak occurs after the secondary eclipse.
For $\tau_{\rm rad}=5~{\rm days}$, the shape of the light curve is nearly the same between planets with $\theta$ (Figure \ref{fig:lcurve_circ}) and ${180}^{\circ}-\theta$ (Figure \ref{fig:lcurve_retro}).
This is because the geometrical effect shown in panel (F) of Figure \ref{fig:lcurve_map} disappears for a planet with a long radiative timescale and a longitudinally homogenized height field (see Figure \ref{fig:regimes}).
In this regime, the shape of the light curve is largely determined by the flux from the poles, which only depends on the subobserver latitude $phi_{\rm obs}$ and $\Lambda$.
Since the subobserver latitude for the case with $\theta$ is identical to that with ${180}^{\circ}-\theta$ (see Equation \ref{eq:phi_obs}), the two planets produce nearly the same light curves for the same $\Lambda$.

We summarize the peak offset calculated from our 2D simulations, as well as the prediction of our analytical theory for tilted planets with various radiative timescales, obliquities, and viewing geometries, in Figure \ref{fig:appendix}.
As seen in Figure \ref{fig:appendix}, our analytical theory well captures the general trends of the peak offset of the light curves of tilted planets.
For short radiative timescales (left panel of Figure \ref{fig:appendix}), the phase of the flux peak is mainly controlled by the shifted hot spot. The flux peak tends to occur near the secondary eclipse as the obliquity approaches $\theta={90}^{\circ}$ and intrinsically occurs after the secondary eclipse for planets with retrograde rotation ($\theta>{90}^{\circ}$).
For long radiative timescales (right panel of Figure \ref{fig:appendix}), the phase of the flux peak is mainly determined by the time lag of the polar heating behind to the solstice $\Lambda$, but the peak offsets for planets with small obliquities  ($\theta \la {18}^{\circ}$) are still influenced by the projected hot spot.

In most cases, our analytical predictions agree very well with the simulation results.
The analytical predictions for $\tau_{\rm rad}=0.1~{\rm day}$ and $\Lambda=\pm {45}^{\circ}$ deviate little from the numerical results. 
This is probably caused by the fact that the analytical model currently ignores the meridional heat transport for a time evolution of the height field at the pole.
The left panel of Figure \ref{fig:appendix} shows that the light curves without the polar flux (dotted lines) underestimate the magnitude of the peak offset, while those with polar flux produce the peak close to the orbital phase of the summer solstice.
This means that the flux from the polar region is overestimated.
The light curves for $\tau_{\rm rad}=5~{\rm days}$ and $\Lambda=\pm{90}^{\circ}$ are also completely different from the numerical results.
In this specific case, the emergent fluxes from the hot spot and the polar region are not important, and the light curves show double flux peaks (see the left middle panel of Figure \ref{fig:lcurve_circ}).
But also note that the light curves are almost flat in these cases.

In summary, for planets in circular orbits, nontilted planets always exhibit a positive peak offset, while tilted planets potentially exhibit a negative peak offset if the solstice takes place after the secondary eclipse and/or the planet is retrograde rotating.
For hot planets with a short radiative timescale in regime (I), the planets with retrograde rotation produce the flux peak after the secondary eclipse because of the geometrical effect. 
Therefore, the negative peak offset potentially indicates $\theta \geq{90}^{\circ}$ for planets with a short radiative timescale.
For relatively cold planets with a long radiative timescale in regimes (II) and (III), the peak of the light curve occurs after the secondary eclipse because of the polar heating at the solstice.
Since the polar heating occurs for $\theta \geq {18}^{\circ}$, one can use the negative peak offset as a diagnosis of high obliquity $\theta \geq{18}^{\circ}$.
However, it should be pointed out that the negative peak offset can also be produced by other mechanisms, for example, the presence of clouds \citep{Demory+13,Oreshenko+16,Parmentier+16}, a westward shift of the hot spot caused by a rotation slower than the orbital motion \citep{Rauscher&Kempton14}, trapped Rossby waves for slow substellar motion \citep{Penn&Vallis17,Penn&Vallis18}, and time-varying winds caused by magnetic fields \citep{Rogers17}.
Therefore, one must be cautious when interpreting the negative peak offset seen in the light curves in future observations.

%%%%%%%%%%%%%%%%%%%%%%%%%
\subsection{Thermal Light Curves of ET Planets}\label{sec:lcurve_ET}
\begin{figure*}[t]
\centering
\includegraphics[clip, width=\hsize]{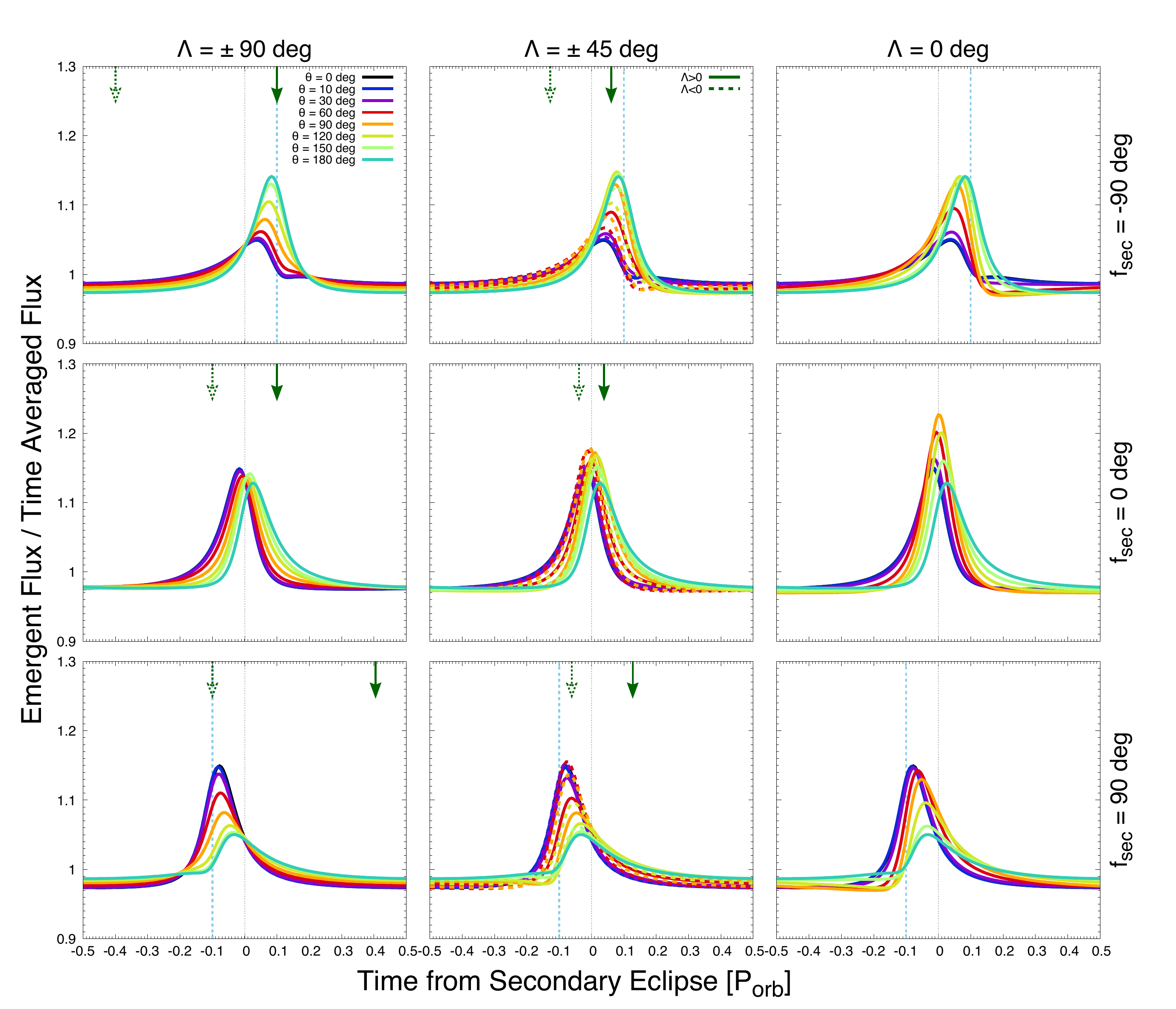}
\caption{
Thermal light curves of ET planets for different $\theta$, $f_{\rm sec}$, and $\Lambda$. The radiative timescale and eccentricity are $\tau_{\rm rad}=0.1~{\rm day}$ and $e=0.5$. Each axis is the same as in Figure \ref{fig:lcurve_circ}. The rows, from top to bottom, exhibit the light curves for the viewing geometry of $f_{\rm sec}=-{90}^{\circ}$, $0^{\circ}$, and ${90}^{\circ}$, respectively. 
The columns, from left to right, show the cases of $\Lambda=\pm{90}^{\circ}$, $\pm{45}^{\circ}$, and ${0}^{\circ}$, respectively.
The different colored lines are the light curves for planets with different obliquity.
The blue dotted lines denote the time of periapse passage.
The green filled and open arrows denote the time of summer solstice that occurs before ($\Lambda<0$) and after ($\Lambda>0$) the secondary eclipse, respectively.
}
\label{fig:lcurve_ET05}
\end{figure*}

\begin{figure*}[t]
\centering
\includegraphics[clip, width=0.9\hsize]{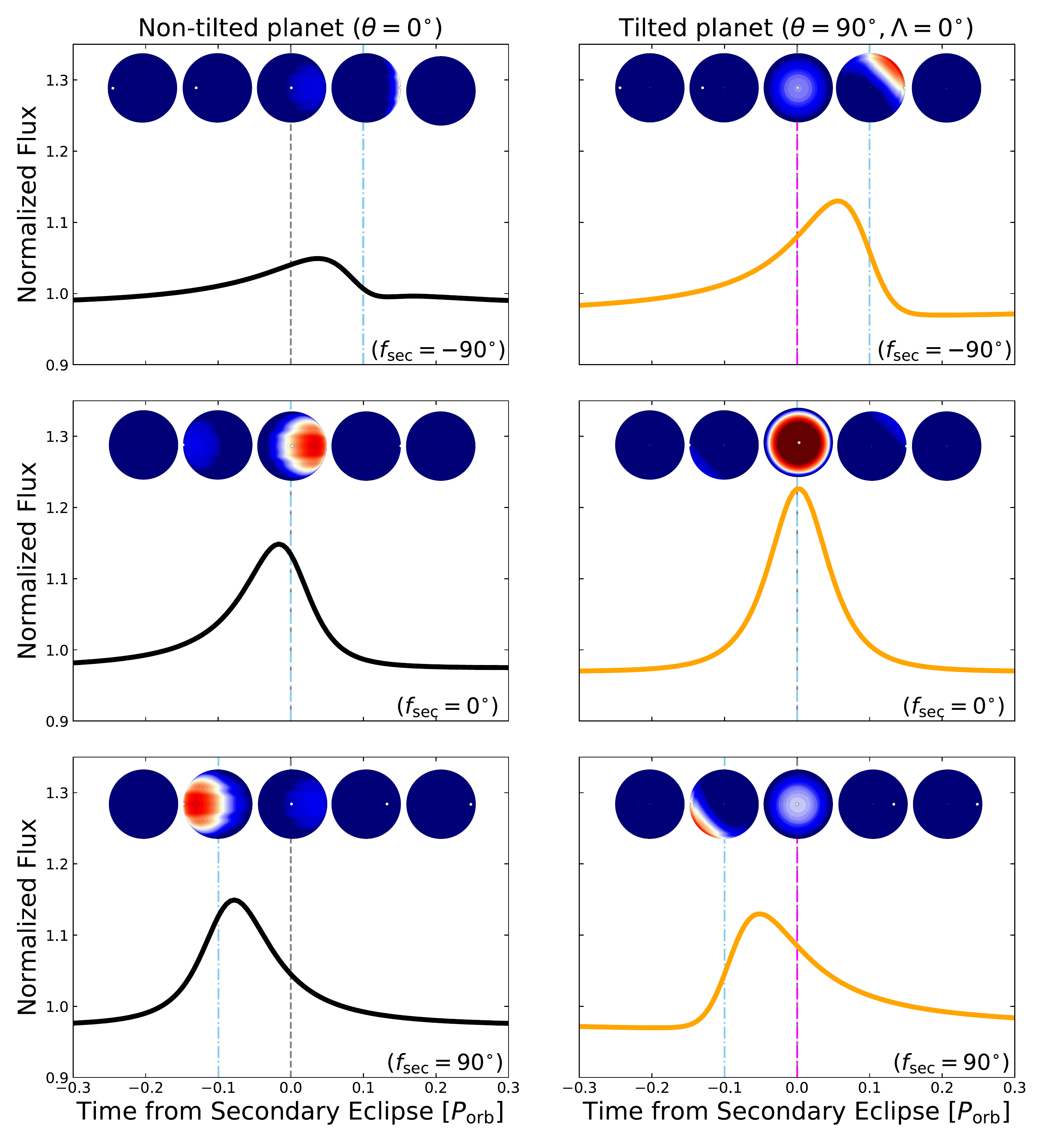}
\caption{Light curves with height fields on the visible hemispheres. The radiative timescale and eccentricity are set as $\tau_{\rm rad}=0.1~{\rm day}$ and $e=0.5$, respectively. The left column shows the light curves and the height field maps for $\theta=0^{\circ}$, while the right column shows those for $\theta={90}^{\circ}$ and the viewing geometry is set to $\Lambda=0$. 
The rows, from top to bottom, show the light curves and the height field maps for $f_{\rm sec}=-{90}^{\circ}$, $0^{\circ}$, and ${90}^{\circ}$, respectively. 
The snapshot of the height field on the visible hemisphere is taken at $t=-0.2$, $-0.1$, $0$, $0.1$, and $0.2P_{\rm orb}$, respectively, where $t=0$ is set to the secondary eclipse timing. The times of the secondary eclipse and the periapse passage are denoted as the gray dotted lines and the blue dash-dotted lines, respectively.
}
\label{fig:phase_t05}
\end{figure*}

\begin{figure*}[t]
\centering
\includegraphics[clip, width=\hsize]{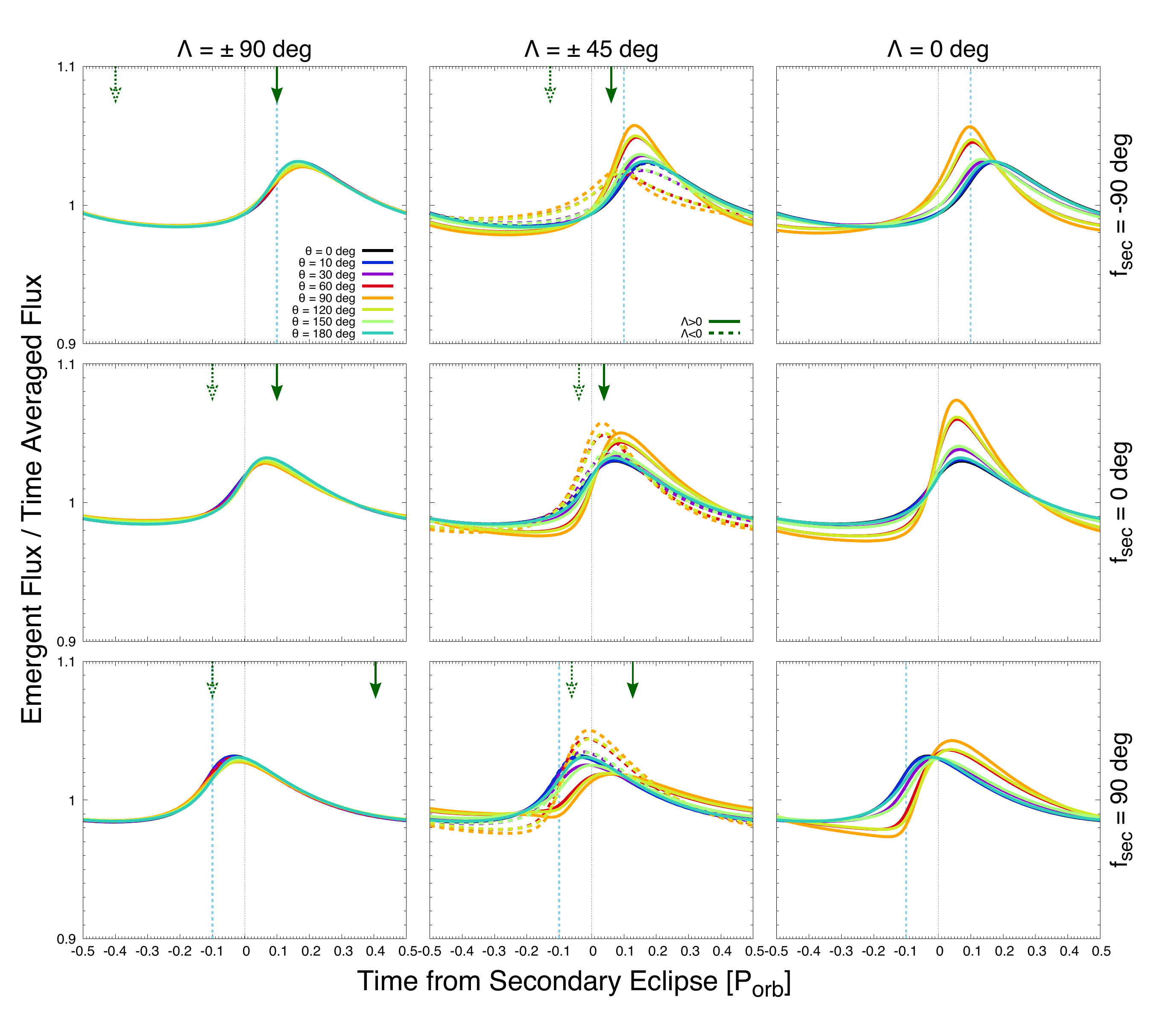}
\caption{
Same as Figure \ref{fig:lcurve_ET05}, but for $\tau_{\rm rad}=5~{\rm days}$.
}
\label{fig:lcurve_ET5}
\end{figure*}

\begin{figure*}[t]
\centering
\includegraphics[clip, width=0.9\hsize]{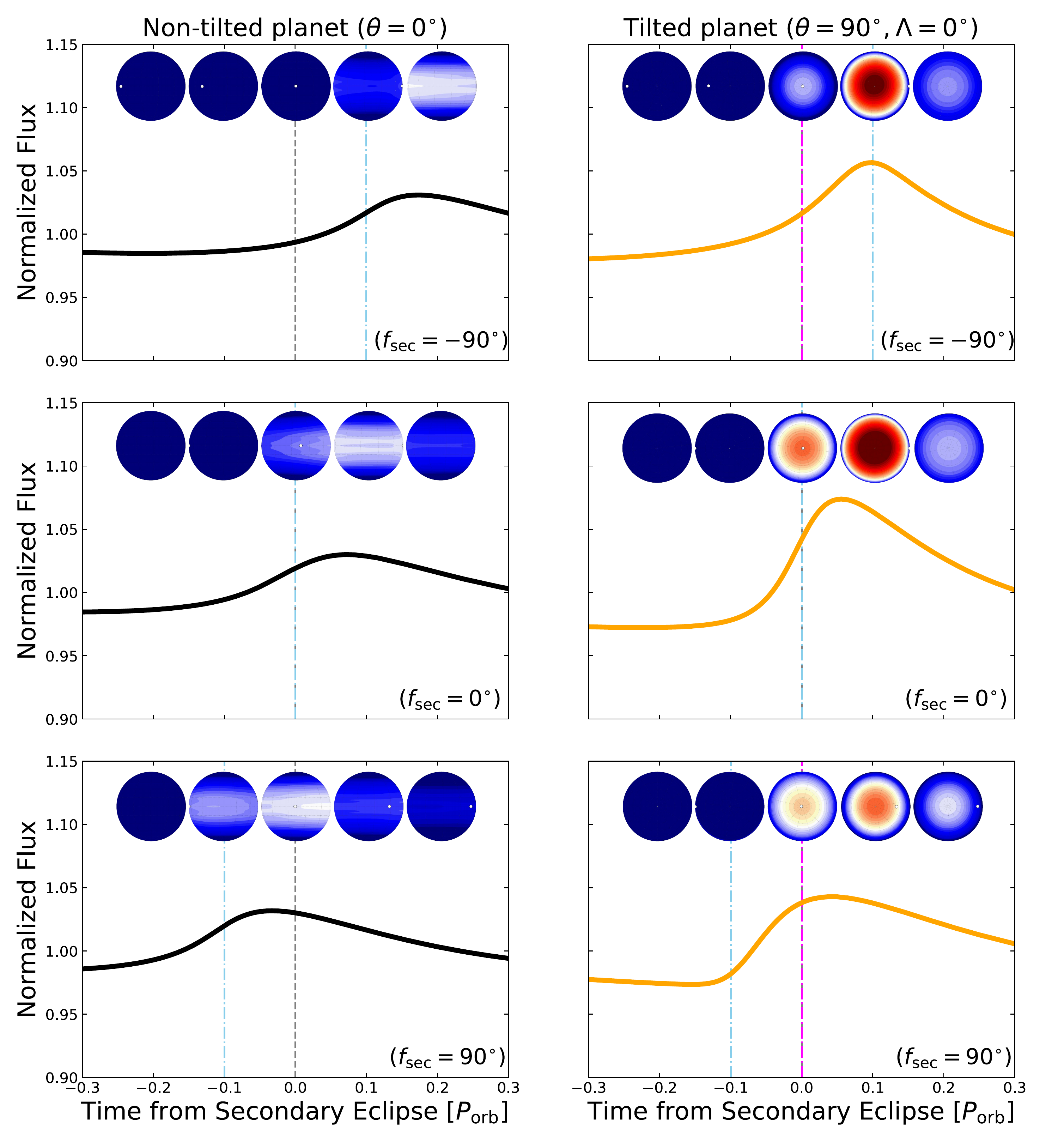}
\caption{Same as Figure \ref{fig:phase_t05}, but for $\tau_{\rm rad}=5~{\rm days}$.
}
\label{fig:phase_t5}
\end{figure*}

%\begin{figure*}[t]
For a planet with a nonzero eccentricity, the shape of the thermal light curve depends not only on eccentricity $e$, obliquity $\theta$, and $\Lambda$ but also on the true anomaly of the secondary eclipse $f_{\rm sec}$.
The situation could be highly complicated if we discuss every possible geometry.
In this study, for simplicity, we follow a similar discussion in \citet{Kataria+13} which also examined the light curves for eccentric planets.
We only discuss the geometries of $f_{\rm sec}=0^{\circ}$ and $\pm{90}^{\circ}$, as shown in Figure \ref{fig:geometry}.
Figure \ref{fig:lcurve_ET05} and \ref{fig:lcurve_ET5} show the light curves of ET planets for a variety of $\theta$ and $\Lambda$ for $\tau_{\rm rad}=0.1$ and $5~{\rm days}$.
The light curves for $\tau_{\rm rad}=100~{\rm days}$ are nearly flat, so we omit them in this section.

When the radiative timescale is short, the eccentricity effect tends to produce a sharp peak near the periapse, where the planet undergoes intense irradiation from the central star. 
Generally speaking, eccentricity leads to the shape of the light curve that is narrow around at the periapse because the time duration for orbiting around the periapse is very short.
For $\tau_{\rm rad}=0.1~{\rm day}$ (Figure \ref{fig:lcurve_ET05}), one can see that the light curves for both tilted and nontilted planets are peaked at around the periapse regardless of the position of the solstice.
This is because the magnitude of the height field is primarily determined by the equilibrium value in the limit of the short radiative timescale.
If the solstice does not take place at the periapse, the height field will be maximized at the periapse and produce the flux peak near the periapse instead of the solstice in the light curve (Figure \ref{fig:lcurve_ET05}).

As noted, for eccentric planets, the shape of the light curve depends on the viewing geometry.
To better explain the effects of the viewing geometry, we show light curves and height fields on the visible hemisphere at an orbital phase around the secondary eclipse for $f_{\rm sec}={0}^{\circ}$ and $\pm{90}^{\circ}$ in Figure \ref{fig:phase_t05}.
For nontilted planets with a short $\tau_{\rm rad}$ (left columns of Figure \ref{fig:phase_t05}), the observed flux peak depends on both the viewing geometry and the hot-spot displacement.
If the periapse passage occurs before the secondary eclipse (for example, $f_{\rm sec}={90}^{\circ}$; left panel of Figure \ref{fig:geometry}), the strong emergent flux can be observed at around the periapse because the hot spot is displaced toward the observer (bottom left panel of Figure \ref{fig:phase_t05}).
By contrast, if the periapse passage occurs after the secondary eclipse (for example, $f_{\rm sec}={-90}^{\circ}$; right panel of Figure \ref{fig:geometry}), the hot spot is displaced away from the observer at the periapse, and the peak flux is smaller than the former case with $f_{\rm sec}={90}^{\circ}$ (top left panel of Figure \ref{fig:phase_t05}).
If the secondary eclipse occurs right at the periapse (middle panel of Figure \ref{fig:geometry}), the strong flux peak occurs slightly before the secondary eclipse because of the eastward displacement of the hot spot (middle left panel of Figure \ref{fig:phase_t05}).
These behaviors are seen in the light curves in Figure \ref{fig:lcurve_ET05}.
Our results are also qualitatively consistent with the light curves from the 3D simulations of planets in eccentric orbits in \citet{Kataria+13} which showed that the peak of the light curve is relatively weak for $f_{\rm sec}={-90}^{\circ}$ ($\omega={360}^{\circ}$ in their context).

The peak offset is insensitive to the obliquity because the emergent flux is largely controlled by the intense heating at the periapse; however, an amplitude of the light curve is drastically influenced by the obliquity.
For example, in the case of $f_{\rm sec}=-{90}^{\circ}$ (top row of Figure \ref{fig:lcurve_ET05}), a tilted planet exhibits a light curve with a large peak as the obliquity approaches $\theta={90}^{\circ}$.
The reason is that, in contrast to the nontilted planets, where the hot spot is displaced away from the observer in this geometry, the emergent flux is dominated by the flux from the heated pole that is still visible at the periapse (see top right panel of Figure \ref{fig:phase_t05}).
For obliquity $\theta>{90}^{\circ}$, the hot spot is displaced toward the observer at the periapse because of the retrograde rotation (Section \ref{sec:lcurve_T}), leading to a large peak in the light curve.
On the other hand, in the case of $f_{\rm sec}={90}^{\circ}$ (bottom row of Figure \ref{fig:lcurve_ET05}), the hot spots on highly tilted planets are not displaced toward the observer at the periapse (see bottom right panel of Figure \ref{fig:phase_t05}).
Therefore, the flux peak occurs closer to the secondary eclipse where the equilibrium height field is smaller than that for the periapse.
Specifically, for obliquity $\theta>{90}^{\circ}$, the hot spot is displaced away from the observer as in nontilted planets for $f_{\rm sec}=-{90}^{\circ}$.
As a result, for $f_{\rm sec}={90}^{\circ}$, the higher $\theta$ case has a smaller peak amplitude in the light curve.
Therefore, for eccentric planets with a short radiative timescale in regime (I), it would be difficult to infer the planetary obliquity from the peak offset.
Alternatively, a light curve with an abnormally large or small peak might imply a high obliquity of the planet in the geometry where the secondary eclipse occurs before ($f_{\rm sec}=-{90}^{\circ}$) and after ($f_{\rm sec}={90}^{\circ}$) the periapse passage, respectively.

As the radiative timescale increases, the peak of the light curve lags behind the planet periapse passage, as seen in the case of $\tau_{\rm rad}=5~{\rm days}$ (Figure \ref{fig:lcurve_ET5}). 
Figure \ref{fig:phase_t5} shows the light curves and the height fields on the visible hemisphere around the secondary eclipse for $\tau_{\rm rad}=5~{\rm days}$.
For nontilted planets, the light curves are always peaked a little behind the periapse passage due to a delayed response of the height field to the stellar heating (left column of Figure \ref{fig:phase_t5}). 
In contrast to the cases of $\tau_{\rm rad}=0.1~{\rm day}$, the shape of the light curve depends less on the viewing geometry, since the height field is nearly homogenized in longitude.
This is why the shapes of the light curves of nontilted planets at different viewing geometries look roughly similar, although the phase of the flux peak also depends on $f_{\rm sec}$.

For tilted planets, the light curves for $\theta={10}^{\circ}$ and ${180}^{\circ}$ are superposed on those for nontilted planets, while the light curves for $\theta = {30}^{\circ}$, ${60}^{\circ}$, ${90}^{\circ}$, ${120}^{\circ}$, and ${150}^{\circ}$ look different.
In addition, light curves for $\theta={30}^{\circ}$ and ${60}^{\circ}$ are almost superposed on those for $\theta={150}^{\circ}$ and ${120}^{\circ}$, respectively.
This behavior is similar to the cases of circular-orbit planets (Section \ref{sec:lcurve_T}).
This indicates that the shape of the light curve is influenced by the polar heating that occurs for obliquity $\theta>{18}^{\circ}$.
However, for $\Lambda=\pm{90}^{\circ}$, the light curves for all obliquities are all superposed on each other because the flux from the poles is negligible in this geometry.
In other words, the light curves of the tilted planets with equinox at the secondary eclipse are indistinguishable from those of the nontilted planets.

For the ET planets with a long radiative timescale, the peak offset of the light curve from the secondary eclipse is more influenced by the obliquity than the planets with short radiative timescales.
For example, in the cases of $\Lambda = {45}^{\circ}$ (solid lines in the middle column of Figure \ref{fig:lcurve_ET5}), the phase of the flux peak is shifted toward the phase of the solstice from the periapse passage.
As obliquity increases, the peak flux is enhanced if the solstice is close to the periapse, as in the case of $f_{\rm sec}=-{90}^{\circ}$ and $0^{\circ}$ (top and middle rows), and is weaken if the solstice is far away from the periapse as in the case of $f_{\rm sec}={90}^{\circ}$ (bottom row).
Take $\Lambda={0}^{\circ}$ as a clearer example (right column); since there is a time delay in the response of the height field to the stellar heating, the flux peak shows a time lag behind the solstice phase (see right column of Figure \ref{fig:phase_t5}).
In summary, when the radiative timescale is as long as classified into regime (II) and (III), the peak offset of the tilted planet is substantially controlled by the solstice phase, while the nontilted planets produce the flux peak around the periapse\footnote{Note that in some architecture, the solstice will also occur around the periapse.}.

As shown above, eccentricity significantly affects the shape of the light curve.
Does it help us to infer the obliquities of eccentric planets?
Here we emphasize that the eccentricity $e$ and $f_{\rm sec}$ are a{\it priori known} parameters from observations, while the obliquity $\theta$ and $f_{\rm sol}$ are a {\it priori unknown} parameters.
If the radiative timescale is significantly short, the flux peak is strongly restricted at around the periapse, where the irradiation is maximum.
In that case, the peak offset depends less on obliquity (see Figure \ref{fig:lcurve_ET05}).
Therefore, in terms of peak offset, it seems difficult to distinguish tilted planets from non-tilted planets, although obliquity significantly affects the peak amplitude that might offer clues to infer the obliquity.
On the other hand, planets with relatively long radiative timescales (in regimes II and III) are better candidates for retrieving the information on the obliquity using the peak offset.
This is because the shape of the light curve easily deviates from that of nontilted planets once the obliquity (use ${180}^{\circ}-\theta$ for $\theta>{90}^{\circ}$) exceeds ${18}^{\circ}$ and the polar heating occurs.

Specifically, we suggest that the observational geometry in which the secondary eclipse takes place after the periapse (i.e., $f_{\rm sec}>0^{\circ}$, for example) offers a good opportunity to infer the obliquity.
In that geometry, the light curve of a nontilted planet tends to exhibit a positive peak offset (i.e., flux peak before the secondary eclipse) because the flux peak is controlled by the heating at the periapse (Figures \ref{fig:lcurve_ET5} and Figure \ref{fig:phase_t5}), although it would also depend on the radiative timescale.
On the other hand, the light curve of a tilted planet potentially exhibits a negative peak offset (i.e., flux peak after the secondary eclipse) because the emergent flux is substantially influenced by the obliquity and the orbital phase of the solstice (the cases of $\Lambda=0^{\circ}$ and ${45}^{\circ}$ in Figure \ref{fig:lcurve_ET5}, for example).
Summarizing, a negative peak offset might indicate a nonzero obliquity (at least, $\theta>{18}^{\circ}$) if the planet is orbiting where the periapse takes place before the secondary eclipse ($f_{\rm sec}>0$).

\begin{figure}[t]
\centering 
\includegraphics[clip, width=\hsize]{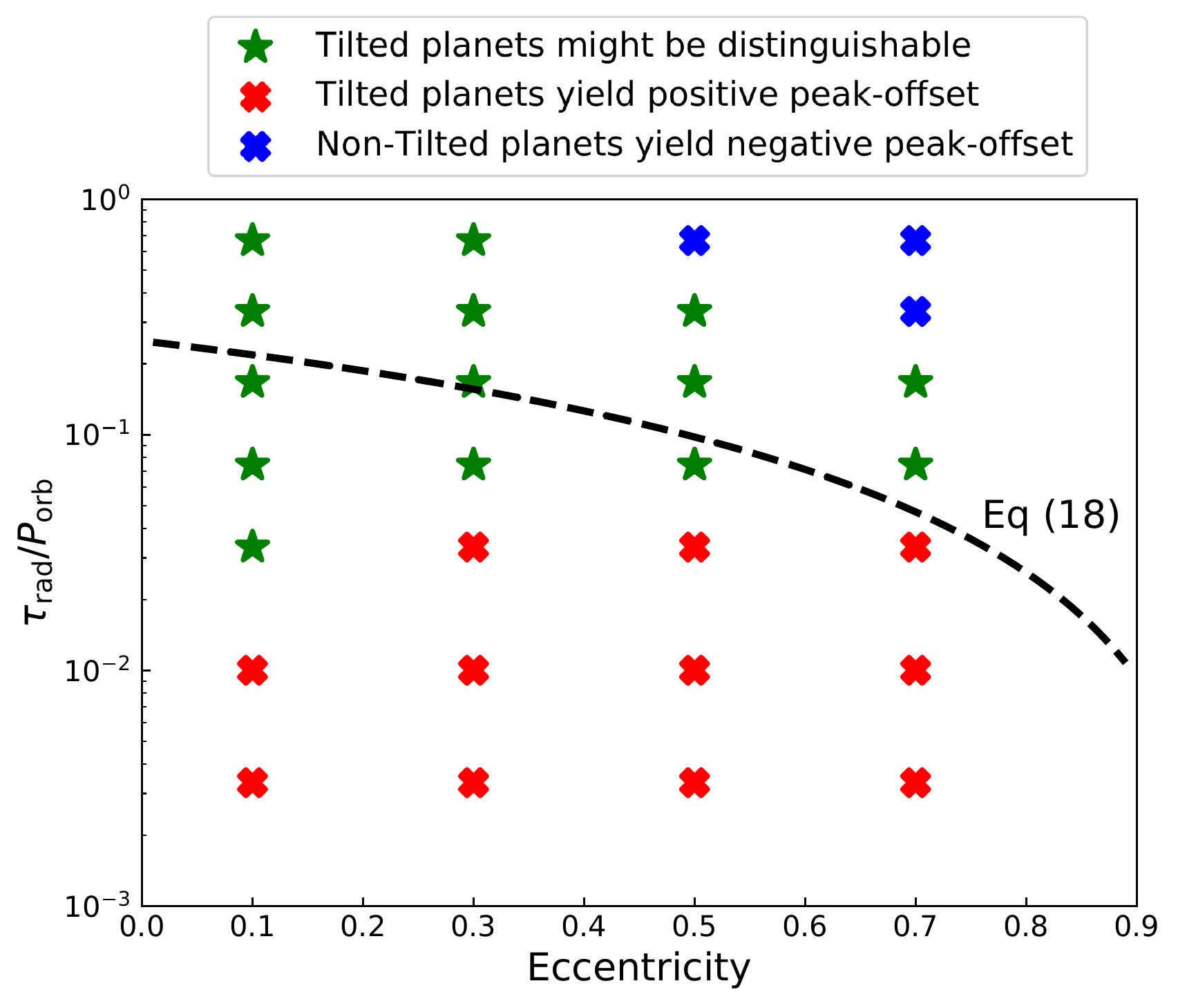}
\caption{Critical radiative timescales (dashed line; Equation \ref{eq:trad_max}) as a function of eccentricity. We assume a viewing geometry of $f_{\rm sec}={90}^{\circ}$. The green stars indicate the parameters for which tilted planets yield a negative peak offset while nontilted planets yield a positive peak offset, confirmed by the light curves from our shallow-water simulations. The red and blue crosses indicate the parameters for which tilted planets yield a positive peak offset and nontilted planets yield a negative peak offset, respectively. The simulations were carried out for $f_{\rm sec}=f_{\rm sol}=\theta={90}^{\circ}$.
}
\label{fig:trad_sweet}
\end{figure}

What radiative timescale is better to infer the obliquity of eccentric planets in that geometry?
If the radiative timescale is too short, a tilted planet produces the flux peak before the secondary eclipse.
Thus, a longer radiative timescale is preferred.
On the other hand, if the radiative timescale is too long, nontilted planets also show the flux peak after the secondary eclipse because of the time delay of the height field response from heating at the periapse.
Therefore, to use the negative peak offset as a diagnosis of the nonzero obliquity, the radiative timescale should not significantly exceed a critical timescale.
Assuming a nontilted planet produces the flux peak after the periapse passage by a time lag of ${\sim}\tau_{\rm rad}$, the critical timescale may be given as the duration a planet takes to travel from the periapse to the secondary eclipse.
The time duration of planet traveling can be calculated from the Kepler equation, given by \citep{Murray&Dermott99}
\begin{equation}\label{eq:Kepler}
\frac{dE}{dt}=\frac{2\pi}{P_{\rm orb}}\frac{1}{1-e\cos{E}},
\end{equation}
where $E$ is the eccentric anomaly associated with the true anomaly $f$ as
\begin{equation}\label{eq:Eccentric_Anomaly}
\tan{\frac{f}{2}}=\sqrt{\frac{1+e}{1-e}}\tan{\frac{E}{2}}.
\end{equation}
Integrating Equation \eqref{eq:Kepler} from $f=0$ to $f_{\rm sec}$ with Equation \eqref{eq:Eccentric_Anomaly}, the critical radiative timescale is given by
\begin{equation}\label{eq:trad_max}
\tau_{\rm crit} = \frac{P_{\rm orb}}{2\pi}\left[ 2{\rm tan}^{-1}\left( \sqrt{\frac{1-e}{1+e}}\tan{\frac{f_{\rm sec}}{2}}\right)-\frac{e\sqrt{1-e^2}\sin{f_{\rm sec}}}{1+e\cos{f_{\rm sec}}} \right]
\end{equation}
Figure \ref{fig:trad_sweet} shows the critical radiative timescale normalized by the orbital period (dashed line) and results from the light curves simulated by the shallow-water model (stars and crosses).
We find that tilted planets produce a negative peak offset, while nontilted planets produce a positive peak offset for the radiative timescale close to ${\sim}\tau_{\rm crit}$.
Tilted planets are still distinguishable for a radiative timescale longer than the critical timescale because nontilted planets produce the flux peak after the periapse passage with a time lag smaller than $\tau_{\rm rad}$ (for example, see Figure \ref{fig:lcurve_ET5}).
This is due to the fact that the incoming stellar flux rapidly decreases as a planet moves far away from the periapse.
The critical radiative timescale is roughly ${\sim}0.1 P_{\rm orb}$ for $f_{\rm sec}={90}^{\circ}$.

%%%%%%%%%%%%%%%%%%%%%%%
%%%%%%%%%%%%%%%%%%%%%%%%%%%%%%%%%%

We now attempt to evaluate the temperature range corresponding to the critical radiative timescale, although in a very crude way.
The radiative timescale is roughly evaluated as \citep{ShowmanGuillot02}
\begin{equation}\label{eq:tau_rad}
\tau_{\rm rad}\sim \frac{P_{\rm ph}}{g}\frac{c_{\rm p}}{4\sigma T^3},
\end{equation}
where $P_{\rm ph}$ is the photospheric temperature, $c_{\rm p}$ is the specific heat of the atmosphere, and $\sigma$ is the Stefan-Boltzmann constant.
If we assume $P_{\rm orb}={30}~{\rm days}$, $c_{\rm p}=1.3\times{10}^{4}~{\rm J~{kg}^{-1}~{K}^{-1}}$, $g=21~{\rm m~{s}^{-2}}$, and $P_{\rm ph}=250$--${670}~{\rm mbar}$ \citep{Showman+15,Rauscher17}, a critical radiative timescale of ${\sim}3~{\rm days}$ corresponds to a temperature of ${\sim}600$--${900}~{\rm K}$.
This suggests that eccentric Jupiter-size planets with equilibrium temperatures of ${\sim}600$--$900~{\rm K}$ might be good candidates to infer nonzero obliquities in future observations.
However, it should be noted that an actual radiative timescale is determined by complex atmospheric properties, such as chemical and thermal structures \citep[e.g.,][]{Li+18}.
Further studies with realistic radiative transfer will be needed to revisit the estimate here.

%\subsection{Influences of Clouds}
%Although tilted planets potentially cause the negative peak-offset, it should be noted that other mechanisms also produce the negative peak-offset as well, which is not ruled out in this study.
%\citet{Rauscher&Kempton14} showed that a non-tilted planet produces the flux peak after the secondary eclipse if the planetary day is longer than the planetary year.
%The negative peak-offset might be also caused by the interaction between the atmosphere and the magnetic field for very %hot planets where a significant thermal ionization of alkali metals takes place \citep{Rogers17}.

%}

%%%%%%%%%%%%%%%%%%%%%%%%%%%%%%%%%%
%%%%%%%%%%%%%%%%%%%%%%%%%%%%%%%%%%
\section{Summary and Discussion}\label{sec:summary}
%%%%%%%%%%%%%%%%%%%%%%%%%%%%%%%%%%
We have investigated the thermal light curves of ET planets for a variety of the radiative timescales, obliquities, orientations of rotation axis, eccentricities, and viewing geometries using the results of the shallow-water simulations presented in \citet{Ohno&Zhang19}.
We have also achieved an analytical theory of thermal light curves for tilted planets for arbitrary radiative timescale, obliquity, and the viewing geometry (Section \ref{sec:lcurve_analy}, Appendix \ref{sec:appendix}).
We discussed how the radiative timescale, obliquity, and eccentricity affect the shape of the light curves and suggested the diagnosis to infer exoplanetary obliquities.
Our findings are summarized as follows.

(1) The shape of the thermal light curve is significantly influenced by the planetary obliquity (Section \ref{sec:lcurve_T}).
For tilted planets with a short radiative timescale, as in regime (I), the peak offset is determined by the hot spot projected onto the orbital plane.
Because of the geometrical effect, tilted planets with retrograde rotation ($\theta>{90}^{\circ}$) produce the flux peak after the secondary eclipse, which is significantly different from nontilted planets.

(2) For tilted planets with a long radiative timescale, as in regimes (II) and (III), the peak of the light curve is largely controlled by the flux from the heated pole if the obliquity is $\theta>{18}^{\circ}$.
Since the polar flux is maximized at around the summer solstice, tilted planets exhibit the flux peak after the secondary eclipse if the solstice takes place after the secondary eclipse.

(3) Our analytical theory of thermal light curves also well explains the basic behaviors observed in numerical light curves.
In summary, a negative peak offset (the peak after the secondary eclipse) potentially implies the planetary obliquity $\theta>{90}^{\circ}$ if the radiative timescale is short and $\theta>{18}^{\circ}$ if the radiative timescale is long, although some other possibilities cannot be ruled out.
%{\bf We have also constructed an analytical thermal light curve for tilted planets in circular orbit (Section \ref{sec:lcurve_analy}, Appendix \ref{sec:appendix}), and found that the analytical model well explains the basic behaviors observed in numerical light curves.}

(4) The shape of the light curve is also significantly influenced by eccentricity (Section \ref{sec:lcurve_ET}).
Because the planet undergoes a more intense heating around the periapse, nontilted planets exhibit a flux peak at around the periapse.
For tilted planets with a short radiative timescale, the light curve also shows a flux peak at around the periapse regardless of the orbital phase of the summer solstice, leading to a degeneracy of the peak offsets between the tilted and nontilted planets.
On the other hand, the obliquity significantly increases or decreases the peak amplitude of the light curve, depending on the viewing geometry, which might offer hints to infer the obliquity if one observes the light curve with an abnormally large or small amplitude.

(5) For tilted planets with a long radiative timescale, the polar heating around the solstice moderately affects the shape of the light curve if the obliquity is higher than ${18}^{\circ}$.
Since both the peak offset and the amplitude are easily influenced by the obliquity on planets with relatively long radiative timescales, they might be better candidates for the observational search for tilted exoplanets.

(6) We suggest that the observational geometry in which the secondary eclipse takes place after the periapse offers a good opportunity to infer the obliquity of a planet in an eccentric orbit (Section \ref{sec:lcurve_ET}).
In this geometry, tilted planets potentially produce the flux peak after the secondary eclipse, whereas nontilted planets produce the peak before the secondary eclipse, when the periapse passage occurs, although it also depends on the radiative timescale.
We suggest that, if the negative peak offset is observed in this geometry, it might imply a tilted planet with an obliquity at least higher than ${18}^{\circ}$.
%{\bf Our simulations indicate that an eccentric planet with a radiative timescale close to the critical timescale (Equation \ref{eq:trad_max}) will be potentially a good candidate to infer a non-zero obliquity in future observations.}

%In this study, we used an idealized 2D shallow water model {\bf to examine a broad parameter space originated by the nature of ET planets.
%Although the idealized model is useful to understand comprehensive behaviors of dynamical regime and observable light curves, the 3D simulations with a realistic radiative transfer is required to quantitively compare the model with observations for a specific object. 
%It may be worth investigating the hypothesis of a non-zero obliquity for planets with negative peak-offset such as Kepler-7 b \citep{Demory+13} and CoRot-2b \citep{Dang+18}.

Although a tilted planet potentially produces a negative peak offset, we should note that there are also other mechanisms causing the negative peak offset in the thermal light curve, as mentioned in Section \ref{sec:lcurve_T}.
It was suggested that, if the planetary day is longer than the planetary year, a nontilted planet also produces a flux peak after the secondary eclipse \citep{Rauscher&Kempton14}.
For very hot planets where a significant ionization of alkali metals takes place, the interaction between atmospheres and magnetic fields potentially causes westward jets, leading to a negative peak offset \citep{Rogers17}. 
If the substellar-point velocity is slower than the Kelvin wave speed, the hot spot is prograde to the substellar point, and the flux peak can take place after the secondary eclipse \citep{Penn&Vallis17,Penn&Vallis18}.
Therefore, one needs to interpret the negative peak offset with caution.

We should also note that the presence of clouds produces a reflected light from the atmosphere that has a significant impact on the shape of the light curve.
Recent photometric observations have detected the negative peak offset for several planets \citep{Demory+13,Angerhausen+15,Esteves+15,Shporer&Hu15,Armstrong+16,Dang+18}, but it probably suggests the presence of clouds in the dayside hemisphere \citep{Oreshenko+16,Parmentier+16}.
The significance of the effects depends on the optical properties and spatial distributions of the clouds, which were recently investigated by cloud microphysical models \citep[e.g.,][]{Helling+08,Lee+15,Gao&Benekke18,Ohno&Okuzumi18,Powell+18,Ormel&Min18} as well as atmospheric circulation models with cloud microphysics \citep{Lee+16,Lines+18}.
Future investigations of atmospheric dynamics coupled with the cloud formation would offer clues to shed light on the diversity of exoplanet light curves and their implications for planetary obliquities.

%Cloud formation on non-synchronized exoplanets have not been well examined yet, so further studies will be necessary to understand how the cloud affects the observable light curves. 
%This would help to distinguish the negative peak offset caused by clouds and non-zero planetary obliquity.

%{\bf As discussed in Section \ref{sec:discussion}, presence of clouds leads to further complexity in atmospheric dynamics and observable light curves.}
%On the other hand, the spatial distribution of clouds is strongly associated with the temperature and dynamical structures \citep{Lee+16,Lines+18}, which are significantly influenced by planetary obliquity.
%Since distinct cloud distributions produce different reflected light curves, it might also provide clues to infer the planetary obliquity.
%The peak offset caused by clouds appreciably depends on the size of cloud particles \citep{Parmentier+16}, which is recently explored by cloud microphysical models %\citep[e.g.,][]{Helling+08,Lee+15,Ohno&Okuzumi18,Powell+18,Gao&Benekke18,Ormel&Min18}.
%Future investigations of atmospheric dynamics coupled with the cloud formation would offer clues to shed light on the diversity of exoplanet light curves and their implications on planetary obliquities.

%%%%%%%%%%%%%%%%%%%%%%%%%%%%%%%%%%%%%%%

\acknowledgments
We thank the anonymous referee for constructive comments and Gongjie Li and Daniel Fabrycky for helpful discussions.
This work was mainly carried out at the Kavli Summer Program in Astrophysics 2016.
We acknowledge Pascal Garaud, Jonathan Fortney, and the entire Kavli scientific organizing committee for thorough support. 
This work was supported by the Kavli Foundation, the National Science Foundation, the Other Worlds Laboratory at UCSC, and University of California Santa Cruz.
K.O. was supported by JSPS KAKENHI Grant Numbers JP15H02065, JP16K17661, JP18J14557, JP18H05438 and the Foundation for Promotion of Astronomy.
X.Z. was supported by NASA Solar System Workings grant NNX16AG08G and NSF Solar and Planetary Research grant AST1740921. 
Most simulations in this study were carried out on the UCSC Hyades supercomputer.

%%%%%%%%%%%%%%%%%%%%%%%%%%%%%%%
%%%%%%%%%%%%%%%%%%%%%%%%%%%%%%%
\appendix
\section{Analytical Theory of the Light Curves for Tilted Planets}\label{sec:appendix}
Here we construct an analytical theory to predict the light curves for tilted planets.
We first present the theory for planets with short radiative timescales, in which the light curve is controlled by the hot spot projected onto the orbital plane, in Appendix \ref{sec:appendix1} and then those for planets with long radiative timescales, in which the flux from the polar region is important, in Appendix \ref{sec:appendix3}.
The theory to evaluate the peak offset for nontilted planets is also presented in Appendix \ref{sec:appendix2}.
Finally, with some assumptions, we construct a universal model of light curves for tilted planets in Appendix \ref{sec:appendix4}.

%%%%%%%%%%%%%%%%%%%%%%%%%%%%%%%
\subsection{Projected Hot Spot for Planets with Short Radiative Timescales}\label{sec:appendix1}
For $\tau_{\rm rad}\ll P_{\rm rot}$, the emergent flux is mainly dominated by the flux from the hot spot displaced from the substellar point.
Therefore, the emergent flux might be diagnosed by the hot-spot vector $\mathbf{r}_{\rm hs}$ projected onto the subobserver point vector $\mathbf{r}_{\rm obs}$, i.e., $\mathbf{r}_{\rm hs}\cdot \mathbf{r}_{\rm obs}$.
The hot-spot vector $\mathbf{r}_{\rm hs}$ is related to the substellar point vector $\mathbf{r}_{\rm ss}$, which is given by \citep[for derivation, see][]{Dobrovolskis09,Dobrovolskis13}
\begin{equation}
\label{eq:sspoint}
\mathbf{r_{\rm ss}}=\left(
    \begin{array}{c}
     (1-\cos{\theta} )\cos{(f-f_{\rm sol})}\sin{\Omega_{\rm rot}t} -\sin{(\Omega_{\rm rot}t-f+f_{\rm sol})} \\
    (1-\cos{\theta} )\cos{(f-f_{\rm sol})}\cos{\Omega_{\rm rot}t} -\cos{(\Omega_{\rm rot}t-f+f_{\rm sol})} \\
    \cos{(f-f_{\rm sol})}\sin{\theta}
    \end{array}
    \right).
\end{equation}
Because the movements of the substellar and subobserver points due to the planetary rotation are the same, one can simplify the problem in a nonrotating framework.
Substituting $\Omega_{\rm rot}=0$ into Equation \eqref{eq:sspoint}, the substellar point movement is expressed by
\begin{equation}
\label{eq:appendix1}
\mathbf{r_{\rm ss}}=\left(
    \begin{array}{c}
     \sin{(f-f_{\rm sol})} \\
    -\cos{\theta}\cos{(f-f_{\rm sol})} \\
    \cos{(f-f_{\rm sol})}\sin{\theta}
    \end{array}
    \right)=
    \left(
    \begin{array}{c}
     \sin{(f-\Lambda)} \\
    -\cos{\theta} \cos{(f-\Lambda)} \\
    \cos{(f-\Lambda)}\sin{\theta}
    \end{array}
    \right),
\end{equation}
where we assume $f_{\rm sec}=0$ for planets in circular orbits and thus $f_{\rm sol}=\Lambda$.
In this context, $f$ expresses the orbital phase from the secondary eclipse.
Since the subobserver point is identical to the substellar point at the secondary eclipse ($f=0$), the subobserver point is given by
\begin{equation}
\label{eq:appendix1}
\mathbf{r_{\rm obs}}=\left(
    \begin{array}{c}
     -\sin{\Lambda} \\
    -\cos{\theta} \cos{\Lambda} \\
    \cos{\Lambda}\sin{\theta}
    \end{array}
    \right).
\end{equation}
On the other hand, the hot spot displaced from the substellar point by the phase shift $\varphi$ (hereafter called the original phase shift) is given by
\begin{equation}
\label{eq:appendix1}
\mathbf{r_{\rm hs}}=
\left(
    \begin{array}{ccc}
      \cos{\varphi} & -\sin{\varphi} & 0 \\
      \sin{\varphi} & \cos{\varphi} & 0 \\
      0 & 0 & 1
    \end{array}
  \right) \mathbf{r}_{\rm ss}
  =
  \left(
    \begin{array}{c}
     \cos{\varphi}\sin{(f-\Lambda)} + \cos{\theta}\sin{\varphi}\cos{(f-\Lambda)} \\
    \sin{\varphi}\sin{(f-\Lambda)} - \cos{\theta}\cos{\varphi}\cos{(f-\Lambda)} \\
    \cos{(f-\Lambda)}\sin{\theta}
    \end{array}
    \right),
\end{equation}
Now the projection factor $\mathcal{P}\equiv\mathbf{r}_{\rm obs}\cdot \mathbf{r}_{\rm hs}$ is obtained as
\begin{eqnarray}\label{eq:projection}
\nonumber
\mathcal{P}&=&\cos{(f-\Lambda)}[({\rm sin}^{2}\theta+\cos{\varphi}~{\rm cos}^{2}\theta)\cos{\Lambda}-\sin{\varphi}\cos{\theta}\sin{\Lambda}]-\sin{(f-\Lambda)}(\cos{\varphi}\sin{\Lambda}+\sin{\varphi}\cos{\theta}\cos{\Lambda})\\
&=&C(\theta,\varphi,\Lambda)\cos{(f-\Lambda+\varphi_{\rm *})},
\end{eqnarray}
where $C(\theta,\varphi,\Lambda)$ is the prefactor controlling the light curve amplitude, given by
\begin{equation}\label{eq:amplitude}
C(\theta,\varphi,\Lambda)=\sqrt{( {\rm sin}^{2}\theta+\cos{\varphi}~{\rm cos}^{2}\theta )^2{\rm cos}^{2}\Lambda-( {\rm sin}^{2}\theta+\cos{\varphi}~{\rm cos}^{2}\theta -\cos{\varphi})\sin{\varphi}\cos{\theta}\sin{2\Lambda}+{\rm cos}^{2}\varphi~{\rm sin}^{2}\Lambda+{\rm sin}^{2}\varphi~{\rm cos}^{2}\theta}
\end{equation}
The projection factor $\mathcal{P}$ is maximized at the orbital phase of $f=\Lambda-\varphi_{\rm *}$, where $\varphi_{\rm *}$ is given by
\begin{equation}\label{eq:appendix2}
\varphi_{\rm *}={\rm tan}^{-1}\left[ \frac{\cos{\varphi}\sin{\Lambda}+\sin{\varphi}\cos{\theta}\cos{\Lambda}}{({\rm sin}^{2}\theta+\cos{\varphi}~{\rm cos}^{2}\theta)\cos{\Lambda}-\sin{\varphi}\cos{\theta}\sin{\Lambda}}\right].
\end{equation}
As seen in Equation \eqref{eq:appendix2}, the peak offset of the tilted planet in regime (I) is determined by a very complex manner.
But if one crudely assumes that the phase shift of the hot spot $\varphi$ is small, which might be valid in the limit of short radiative timescales and weak zonal winds, the problem is greatly simplified. 
In that case, Equation \eqref{eq:appendix2} can be approximated as
\begin{eqnarray}\label{eq:appendix3}
\nonumber
\varphi_{\rm *}&\approx& {\rm tan}^{-1}\left[ \frac{\sin{\Lambda}+\varphi\cos{\theta}\cos{\Lambda}}{\cos{\Lambda}-\varphi\cos{\theta}\sin{\Lambda}}\right]\\
\nonumber
&\approx& {\rm tan}^{-1}\left[ \frac{\sin{(\Lambda+{\rm tan}^{-1}(\varphi \cos{\theta}))}}{ \cos{(\Lambda+{\rm tan}^{-1}(\varphi \cos{\theta}))} }\right]\\
&\approx& \Lambda + \varphi \cos{\theta},
\end{eqnarray}
where we truncate the terms of $\varphi$ higher than the second order.
Equation \eqref{eq:appendix3} indicates that, for a small phase shift of the hot spot, the flux peak occurs at $f=-\varphi \cos{\theta}$ which is just the projection of the original phase shift onto the orbital plane.

Here we show some examples to interpret the light curves for a short radiative timescale in Section \ref{sec:lcurve_T}.
In the simplest case, the projection factor for a nontilted planet ($\theta=0$) is given by
\begin{equation}
\mathcal{P}|_{\rm \theta=0}=\cos{(f+\varphi)},
\end{equation}
where $\mathcal{P}|_{\rm \theta=0^{\circ}}$ is maximized at $f=-\varphi$ and thus the flux peak occurs before the secondary eclipse.
On the other hand, for $\theta={180}^{\circ}$, the projection factor is given by
\begin{equation}
\mathcal{P}|_{\rm \theta={180}^{\circ}}=\cos{(f-\varphi)},
\end{equation}
where $\mathcal{P}|_{\rm \theta={180}^{\circ}}$ is maximized at $f=\varphi$ and thus the flux peak occurs after the secondary eclipse.
For another example, the projection factor for planets in the geometry of $\Lambda={90}^{\circ}$ is given by
\begin{equation}\label{eq:projection90}
\mathcal{P}|_{\rm \Lambda={90}^{\circ}}=\sqrt{{\rm cos}^{2}\varphi+{\rm sin}^{2}\varphi~{\rm cos}^{2}\theta}\cos{(f+\varphi_{\rm 90})},
\end{equation}
where $\varphi_{\rm 90}={\rm tan}^{-1}(\cos{\theta}\tan{\varphi})$ and the flux peak occurs at $f=-\varphi_{\rm 90}$.
On the other hand, the projection factor for $\Lambda=0^{\circ}$ is also given by
\begin{equation}\label{eq:projection0}
\mathcal{P}|_{\rm \Lambda={0}^{\circ}}=\sqrt{({\rm sin}^{2}\theta+\cos{\varphi}~{\rm cos}^{2}\theta)^2+{\rm sin}^{2}\varphi~{\rm cos}^{2}\theta}\cos{(f+\varphi_{\rm 0})},
\end{equation}
where $\varphi_{\rm 0}={\rm tan}^{-1}(\cos{\theta}\sin{\varphi}/({\rm sin}^{2}\theta+\cos{\varphi}~{\rm cos}^{2}\theta))$ and the flux peak occurs at $f=-\varphi_{\rm 0}$.

%%%%%%%%%%%%%%%%%%%%%%%%%%%%%%%
\subsection{Original Phase Shift of the Hot Spot}\label{sec:appendix2}
The original phase shift of the hot spot $\varphi$ can be evaluated from linearized shallow-water equations. 
%{\bf The one-dimensional shallow-water equations at the equator are given by \citep[][]{Penn&Vallis17}
%\begin{equation}
%\label{eq:moment}
%\frac{\partial \mathbf{u}}{\partial t}+u\frac{\partial u}{\partial x}+g\frac{\partial h}{\partial x}= -\frac{\mathbf{v}Q}{h}-\frac{\mathbf{v}}{\tau_{\rm drag}},
%\end{equation}
%\begin{equation}
%\label{eq:mass}
%\frac{\partial h}{\partial t} +\nabla \cdot(\mathbf{v}h) = \frac{h_{\rm eq}(\lambda,\phi,t)-h}{\tau_{\rm rad}} \equiv Q,
%\end{equation}
%}
Without the drag and Coriolis term, a linearized shallow-water equation at the equator can be written as \citep[for the derivation, see][]{Penn&Vallis17}
\begin{equation}\label{eq:appendix4}
\frac{dh'}{d \lambda'}=-\xi h' + \xi \Delta h \cos{\lambda'}\mathcal{H}(\cos{\lambda'}),
\end{equation}
where $\mathcal{H}$ is the Heaviside step function defined as Equation \eqref{eq:Heaviside}, $h'=h-H$ is the difference of the height from the mean value, $\lambda'=\lambda+2\pi t/P_{\rm rot}$ is the longitude from the substellar point moving westward, and $\xi$ is the nondimensional parameter given by
\begin{equation}
\xi = \frac{P_{\rm rot}}{2\pi \tau_{\rm rad}}\left( 1-\frac{gH}{v_{\rm ss}^2}\right)^{-1},
\end{equation}
where $v_{\rm ss}=2\pi R_{\rm p}/P_{\rm rot}$ is the substellar velocity.
Note that the substellar point is assumed to move westward since we have assumed $P_{\rm rot}\ll P_{\rm orb}$.
The analytical solution of Equation \eqref{eq:appendix4} for $-\pi/2<\lambda<\pi/2$ is given by
\begin{equation}\label{eq:height_hs}
h'=\Delta h\frac{\xi}{1+\xi^2}\left( \xi \cos{\lambda'}+\sin{\lambda'}+\frac{e^{(3/2)\pi \xi}+e^{(1/2)\pi \xi}}{e^{2\pi \xi}-1}\exp{(-\xi \lambda')} \right),
\end{equation}
where we adopt a periodic boundary condition at $\lambda'=\pm \pi/2$ and $\pm \pi$ following \citet{Zhang&Showman16}.
The phase shift of the hot spot $\varphi$ can be found as a solution of $dh'/d\lambda'=0$, i.e.,
\begin{equation}\label{eq:appendix_h}
\xi \sin{\varphi}- \cos{\varphi}+\xi \frac{e^{(3/2)\pi \xi}+e^{(1/2)\pi \xi}}{e^{2\pi \xi}-1}\exp{(-\xi \varphi)}=0.
\end{equation}
The solution of this equation cannot be explicitly shown but can be numerically obtained.
In the limit of $\xi \gg 1$, the phase shift is expressed by $\varphi \approx{\rm tan}^{-1}(1/\xi)$.

%%%%%%%%%%%%%%%%%%%%%%%%%%%%%%%
\subsection{Evaluation of the Flux from the Polar Region}\label{sec:appendix3}
%%%%%%%%%%%%%%%
\begin{figure}[t]
\centering
\includegraphics[clip, width=0.5\hsize]{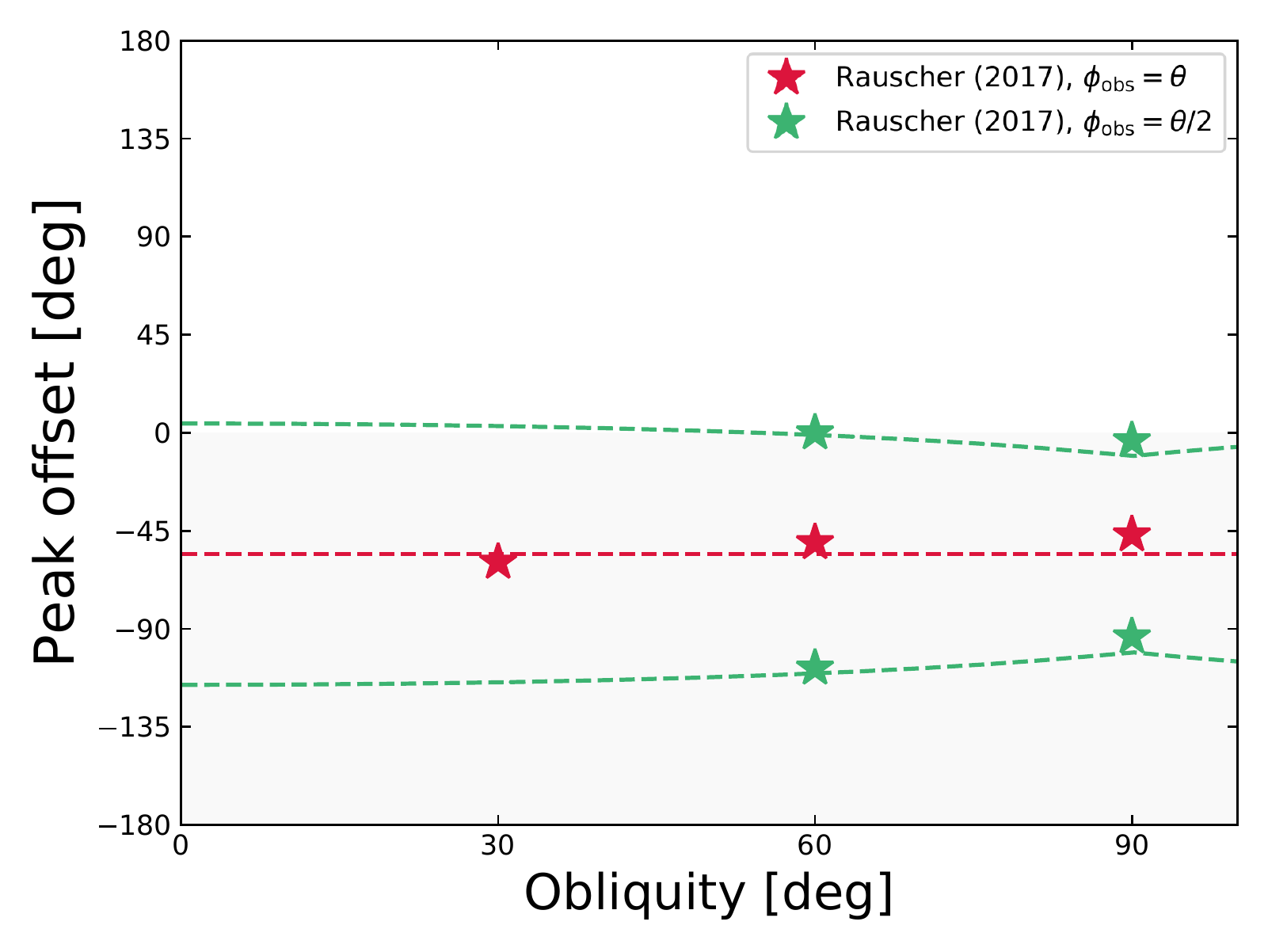}
\caption{Comparison with the peak offset simulated by \citet{Rauscher17}. The stars denote the peak-offset in light curves simulated by \citet{Rauscher17}, where we only plot for light curves with clear peaks. The red and green star-symbols show the peak offset in viewing geometries of $\phi_{\rm obs}=\theta$ and $\theta/2$, respectively. The dotted lines show the prediction of our analytical theory (Equation \ref{eq:fpeak_pole}). We set $P_{\rm rot}={10}~{\rm hours}$, $P_{\rm orb}={10}~{\rm days}$, and $\tau_{\rm rad}=2.3~{\rm days}$ calculated by Equation \eqref{eq:tau_rad} with $P_{\rm ph}=667~{\rm mbar}$ and $T_{\rm eq}=880~{\rm K}$, which are equivalent to the parameters used in \citet{Rauscher17}. 
}
\label{fig:revise1}
\end{figure}
%%%%%%%%%%%%%%%%%

For planets with long radiative timescales of $\tau_{\rm rad}\gg P_{\rm rot}$, the hot spot is less important for the total emergent flux.
When the obliquity is small, as in regime (II), the shape of the light curve is still influenced by the projected hot spot argued in Appendix \ref{sec:appendix1}.
On the other hand, if the planetary obliquity is high, as in regime (III), the total flux is largely dominated by the flux from the polar region.
To evaluate the flux from the polar region, we construct a simple kinematic model of the height field evolution at the pole.
When the meridional heat transport is inefficient, the time evolution of the height field at the pole is expressed by
\begin{equation}\label{eq:height_pole_evolve}
\frac{2\pi}{P_{\rm orb}}\frac{d h'}{df}=\frac{\Delta h \sin{\theta}\cos{(f-\Lambda)}\mathcal{H}(\cos{(f-\Lambda)})-h'}{\tau_{\rm rad}}
\end{equation}
This equation can be solved in the same way as Equation \eqref{eq:appendix4}.
Again adopting the periodic boundary condition at $f=\Lambda \pm \pi/2$ and $\Lambda \pm \pi$, we obtain the height field evolution at the pole,
\begin{equation}\label{eq:height_pole}
h'_{\rm pole}=\Delta h\frac{\psi \sin{\theta}}{1+\psi^2}\left[ \psi \cos{(f-\Lambda)}+\sin{(f-\Lambda)}+\frac{e^{\psi(\Lambda-3\pi/2)}+e^{\psi(\Lambda-2\pi)}}{1-e^{-3\pi \psi/2}}\exp{(-\psi f)} \right],
\end{equation}
where we define
\begin{equation}\label{eq:psi}
\psi \equiv \frac{P_{\rm orb}}{2\pi \tau_{\rm rad}}
\end{equation}
The height field at the pole is maximized when $dh'_{\rm pole}/df=0$ is satisfied, i.e.,
\begin{equation}\label{eq:diff_pole}
\psi \sin{(f-\Lambda)}- \cos{(f-\Lambda)}+\psi \frac{e^{\psi(\Lambda-3\pi/2)}+e^{\psi(\Lambda-2\pi)}}{1-e^{-3\pi \psi/2}}\exp{(-\psi f)}=0.
\end{equation}
Again, one needs to solve this equation numerically. 
But, in the limit of $\psi \gg1$, the phase of the maximum height field is given by
\begin{equation}\label{eq:fpeak_pole}
f=\Lambda+{\rm tan}^{-1}\left(\frac{1}{\psi}\right)\approx \Lambda+\frac{2\pi \tau_{\rm rad}}{P_{\rm orb}}.
\end{equation}
This indicates that the flux from the polar region is maximized after the planet passes the solstice.
Therefore, the phase of the light curve occurs after the solstice passage, which is in agreement with the light curves in regime (III) in Figure \ref{fig:lcurve_circ}.
Equation \eqref{eq:diff_pole} also indicates that the phase of the flux peak from the polar region is independent of the obliquity, which explains why the planets with different obliquities produce the flux peak at the same orbital phase for fixed $\Lambda$ in Figure \ref{fig:lcurve_circ}.
\citet{Rauscher17} assumed that the diurnally averaged insolation, and thus the simulated light curves, should fall into this regime.
Figure \ref{fig:revise1} shows the peak offset retrieved from the light curves in \citet{Rauscher17} and our analytical model (Equation \ref{eq:limit2}). 
The theory presented here excellently matches to the results of \citet{Rauscher17}, implying that the flux from the polar region indeed controls the shape of the light curves in this regime.
%The comparison also supports the importance of flux from the polar region for planets with long radiative timescales.

The light curves for regimes (IV) and (V) can also be understood from the limit of $\psi \ll 1$.
In the limit of $\psi \ll 1$, the height field at the pole is approximated as
\begin{equation}\label{eq:height_pole_annual}
h'_{\rm pole}\approx \Delta h\psi \sin{\theta}\left[ \sin{(f-\Lambda)}+\frac{4+\psi(4\Lambda-7\pi)}{3\pi \psi} \right].
\end{equation}
In this case, the flux peak occurs at the $f=\Lambda+\pi/2$, and this is consistent with the light curves for regimes (IV) and (V) in Figure \ref{fig:lcurve_circ}, although the light curves are nearly flat because the amplitude scales as $\Delta h \psi \sin{\theta}$.

%%%%%%%%%%%%%%%%%%%%%%%%%%%%%%%
\subsection{Universal Formula of Peak Offset for Tilted Planets in Circular Orbits}\label{sec:appendix4}
We now attempt to derive a universal formula of the peak offset for tilted planets.
The transition from the short-$\tau_{\rm rad}$ limit to the long-$\tau_{\rm rad}$ and high-obliquity limit occurs when the flux from the polar region becomes stronger than the flux from the hot spot shifted from the substellar point.
To express the smooth transition of these two limits, we construct a toy model; the height field is excited from the mean value only at the shifted hot spot and the pole.
In this context, the height field is expressed as
\begin{equation}\label{eq:delta}
h=H+h'(\mathbf{r})(\delta(\mathbf{r}-\mathbf{r}_{\rm hs})+\delta(\mathbf{r}-\mathbf{r}_{\rm pole})),
\end{equation}
where $\delta$ is the delta function.
Substituting Equation \eqref{eq:delta} into Equation \eqref{eq:lcurve}, the diagnosis of the emergent flux is given by
\begin{equation}\label{eq:appendix_flux}
F=\int_{\rm 0}^{2\pi} \int_{\rm 0}^{\pi/2} [H+h(\mathbf{r})(\delta(\mathbf{r}-\mathbf{r}_{\rm hs})+\delta(\mathbf{r}-\mathbf{r}_{\rm pole}))](\mathbf{r}\cdot \mathbf{r}_{\rm obs}) \sin{\phi'}d\phi' d\lambda',
\end{equation}
where $\phi'$ is the angle from the subobserver latitude (i.e., $\phi'={\rm cos}^{-1}(\mathbf{r}\cdot \mathbf{r}_{\rm obs})$) and $\lambda'$ is the longitude on the plane perpendicular to the subobserver point vector.
Equation \eqref{eq:appendix_flux} yields
\begin{eqnarray}
\nonumber
F&=&\pi H+h'(\mathbf{r}_{\rm hs}) (\mathbf{r}_{\rm obs}\cdot \mathbf{r}_{\rm hs})  \mathcal{H}(\mathbf{r}_{\rm obs}\cdot \mathbf{r}_{\rm hs})  \sqrt{1-(\mathbf{r}_{\rm obs}\cdot \mathbf{r}_{\rm hs}) ^2}+h'(\mathbf{r}_{\rm pole}) (\mathbf{r}_{\rm obs}\cdot \mathbf{r}_{\rm pole}) \sqrt{1-(\mathbf{r}_{\rm obs}\cdot \mathbf{r}_{\rm pole})^2}\\
&\approx& \pi H +h'(\mathbf{r}_{\rm hs}) \mathcal{P}  \mathcal{H}(\mathcal{P}) +h'(\mathbf{r}_{\rm pole}) \cos{\Lambda}\sin{\theta}.
\end{eqnarray}
The first, second, and third terms express the emergent flux from the mean height field, shifted hot spot, and the pole, respectively.
Inserting Equation \eqref{eq:appendix_h} into Equation \eqref{eq:height_hs} with $\lambda'=\varphi$, the height field at the hot spot is given by
\begin{equation}
h'(\mathbf{r}_{\rm hs})=\Delta h \cos{\varphi}.
\end{equation}
Therefore, using Equation \eqref{eq:projection}, the second term is expressed as
\begin{eqnarray}\label{eq:F_hs}
\nonumber
h'(\mathbf{r}_{\rm hs}) \mathcal{P} &=&\Delta h \cos{\varphi} \{\cos{(f-\Lambda)}[({\rm sin}^{2}\theta+\cos{\varphi}~{\rm cos}^{2}\theta)\cos{\Lambda}-\sin{\varphi}\cos{\theta}\sin{\Lambda}]\\
&&-\sin{(f-\Lambda)}(\cos{\varphi}\sin{\Lambda}+\sin{\varphi}\cos{\theta}\cos{\Lambda})\},
\end{eqnarray}
where we focus on the orbital phase around the flux peak ($\mathcal{P}>0$) and thus $\mathcal{H}(\mathcal{P})=1$.
On the other hand, from Equation \eqref{eq:height_pole}, the flux from the pole is expressed by
\begin{equation}\label{eq:F_pole}
h'(\mathbf{r}_{\rm pole})\cos{\Lambda}\sin{\theta}=\Delta h \frac{\psi \cos{\Lambda}~{\rm sin}^{2}{\theta}}{1+\psi^2}\left[\psi \cos{(f-\Lambda)}+\sin{(f-\Lambda)}+\frac{e^{\psi(\Lambda-3\pi/2)}+e^{\psi(\Lambda-2\pi)}}{1-e^{-3\pi \psi/2}}\exp{(-\psi f)}\right].
\end{equation}
Here we only consider the case of $\psi \gg 1$ and truncate the last term in the bracket of Equation \eqref{eq:F_pole} because the light curve is nearly flat in the limit of $\psi \ll 1$.
From Equations \eqref{eq:F_hs} and \eqref{eq:F_pole}, we obtain the total emergent flux as
\begin{eqnarray}\label{eq:Analytic_Flux}
\nonumber
F(f)&=& \pi H+\Delta h \cos{(f-\Lambda)}\left[\cos{\varphi}(({\rm sin}^{2}\theta+\cos{\varphi}~{\rm cos}^{2}\theta)\cos{\Lambda}-\sin{\varphi}\cos{\theta}\sin{\Lambda})+\frac{\psi^2 \cos{\Lambda}~{\rm sin}^{2}{\theta} }{1+\psi^2}\right]\\
\nonumber
&&~~~~~~ -\Delta h\sin{(f-\Lambda)}\left[\cos{\varphi}(\cos{\varphi}\sin{\Lambda}+\sin{\varphi}\cos{\theta}\cos{\Lambda})-\frac{\psi \cos{\Lambda}~{\rm sin}^{2}{\theta} }{1+\psi^2} \right]\\
&=&\pi H+\Delta hC_{t}(\theta,\varphi,\Lambda,\psi)\cos{[f-\Lambda+\varphi_{\rm peak}(\theta,\varphi,\Lambda,\psi)]},
\end{eqnarray}
where $C_{\rm t}(\theta,\varphi,\Lambda,\psi)$ is the prefactor controlling the light-curve amplitude, given by
\begin{eqnarray}\label{eq:Ct}
\nonumber
C_{\rm t}^2&=&C^2~{\rm cos}^{2}\varphi +\frac{\psi^2~{\rm cos}^{2}\Lambda~{\rm sin}^{4}\theta}{1+\psi^2}\\
& &+\frac{2\psi \cos{\varphi}\cos{\Lambda}~{\rm sin}^{2}\theta}{1+\psi^2}\left[ \psi(({\rm sin}^{2}\theta+\cos{\varphi}~{\rm cos}^{2}\theta)\cos{\Lambda}-\sin{\varphi}\cos{\theta}\sin{\Lambda})-(\cos{\varphi}\sin{\Lambda}+\sin{\varphi}\cos{\theta}\cos{\Lambda}) \right]
\end{eqnarray}
The total flux is maximized at $f_{\rm peak}=\Lambda-\varphi_{\rm peak}$ where
\begin{equation}\label{eq:appendix_full}
\varphi_{\rm peak}={\rm tan}^{-1}\left[ \frac{ (1+\psi^2)(\cos{\varphi}\sin{\Lambda}+\sin{\varphi}\cos{\theta}\cos{\Lambda})\cos{\varphi}-\psi \cos{\Lambda}~{\rm sin}^2{\theta} }{(1+\psi^2)[({\rm sin}^{2}\theta+\cos{\varphi}~{\rm cos}^{2}\theta)\cos{\Lambda}-\sin{\varphi}\cos{\theta}\sin{\Lambda}]\cos{\varphi}+\psi^2\cos{\Lambda}~{\rm sin}^2{\theta}}\right].
\end{equation}
In the limit of short radiative timescales ($\psi \gg 1$) and small obliquities, Equation \eqref{eq:appendix_full} returns to Equation \eqref{eq:appendix2}.
On the other hand, in the limit of radiative timescales longer than the planet day (i.e., $\xi \ll 1$ and thus $\varphi \rightarrow \pi/2$), Equation \eqref{eq:appendix_full} returns to $\varphi_{\rm peak}\approx-{\rm tan}^{-1}(1/\psi)$.
Consequently, the phase of the flux peak can be evaluated by $f_{\rm peak}=\Lambda-\varphi_{\rm peak}$ with Equation \eqref{eq:appendix_full} for arbitrary radiative timescales, obliquities, and viewing geometries.

%\acknowledgments 
%%%%%%%%%%%%%%%%%%%%%%%%%%%%%%%%%%%%%%%

\end{document}